\documentclass[12pt, a4paper]{article}

\usepackage[margin=2.5cm]{geometry}
\usepackage{helvet,times}
\usepackage{bm,textcomp}
\usepackage{amssymb}
\usepackage{amsmath}
\usepackage{graphicx}
\usepackage{subfigure}
\usepackage[percent]{overpic}
\usepackage[table]{xcolor}
\usepackage{marvosym}
\usepackage{appendix}
\usepackage{rotating}
\usepackage[round,authoryear,sort&compress]{natbib}

\newcommand{\envelope}{(\raisebox{-.5pt}{\scalebox{1.45}{\Letter}}\kern+0.1pt)}
\newcommand{\brac}[1]{\left( {#1} \right)}    
\newcommand{\bracs}[1]{\left[ {#1} \right]}   
\newcommand{\bracc}[1]{\left\{ {#1} \right\}} 
\newcommand{\abs}[1]{\left\vert {#1} \right\vert} 

\begin{document}
\title{The mechano-chemistry of cytoskeletal force generation}
\author{
M. Maraldi$^1$, K. Garikipati$^{1, 2}$\footnote{correspondence: krishna@umich.edu} \\
\small{$^1$Department of Mechanical Engineering, University of Michigan, Ann Arbor, Michigan} \\ 
\small{$^2$Department of Mathematics, University of Michigan, Ann Arbor, Michigan}
}
\date{}
\maketitle

\begin{abstract}
In this communication, we propose a model to study the non-equilibrium process by which actin stress fibers develop force in contractile cells.
The emphasis here is on the non-equilibrium thermodynamics, which is necessary to address the mechanics as well as the chemistry of dynamic cell contractility. In this setting we are able to develop a framework that relates (a) the dynamics of force generation within the cell and (b) the cell's response to external stimuli to the chemical processes occurring within the cell, as well as to the mechanics of linkage between the stress fibers, focal adhesions and extra-cellular matrix.
%
\end{abstract}
%
\section{Introduction}\label{intro}
Actin stress fibers can be found in cells such as fibroblasts and muscle cells and play an important role in cell locomotion, cell adhesion and wound healing, due to their ability to generate forces over the entire extent of the cell.
They are constituted by several bundles of actin filaments held together by the actin binding protein $\alpha$-actinin, and others.
Myosin proteins make the actin filaments slide past each other and confer on the stress fiber the ability to generate contractile force \citep{pellegrin:2007}; in fact, it is well-established that a cell's traction force development is associated with stress fiber formation \citep{chrzanowska:1996,ingber:2003}. 
Besides, the kinetics of stress fiber formation and disassembly \citep{pollard:2003} appears itself to be modulated by force: traction force promotes the binding of stress fiber proteins \citep{colombelli:2009,hirata:2008}, which in turn enhances acto-myosin contractile activity, establishing a ``feed-forward'' process for stress fibers growth.

Stress fibers are anchored to the extra-cellular matrix (ECM) through focal adhesions

\noindent\citep{chrzanowska:1996,geiger:2001}.
Focal adhesion proteins including zyxin, paxillin, talin, vinculin and many others, connect integrin (the trans-membrane adhesion protein) to F-actin bundles of stress fibers \citep{geiger:2001,zamir:2001}.
The kinetics of binding/unbinding of focal adhesion proteins is also modulated by force; moreover, it has been shown that focal adhesions exhibit this mechano-sensitivity regardless of whether the force is imposed by acto-myosin contractility of stress fibers, or is externally-imposed and merely transmitted by the stress fibers \citep{balaban:2001,riveline:2001}.

In recent years, a large number of experimental studies have been conducted to investigate the behavior and the properties of the cytoskeletal system: cell traction force \citep{tan:2003}, cell response and differentiation induced by changes in the substrate stiffness \citep{chan:2008,engler:2006},stress fiber orientation parallel to \citep{franke:1984} or away from the direction of cyclic stretching \citep{kaunas:2005}, cell contractility stimulated by treatment with calyculin \citep{peterson:2004} and stress fiber relaxation upon severance by a laser \citep{kumar:2006} among others.
Concurrently, a number of studies concerned with the modeling of the mechano-chemical response of stress fibers have appeared. Some of these studies do not explicitly consider focal adhesions \citep{besser:2011, besser:2007, kaunas:2010, kaunas:2008, kruse:2000, stachowiak:2008, stachowiak:2009}, some others focus solely on modeling the focal adhesions  \citep{shemesh:2005,besser:2006,olberding:2010}, while others do indeed model the coupled system of stress fibers and focal adhesions \citep{deshpande:2006,deshpande:2008,harland:2011,walcott:2010}. 
The diversity of experimental studies, and theoretical models motivated by these experiments, have begun to shed some light on the role of stress fibers and focal adhesions in the development of cell traction.
However, despite these studies there is not yet a complete understanding of how the dynamic mechano-chemistry of stress fibers affects that of focal adhesions, and vice-versa. 

In this paper, we propose a model able to describe the interaction between stress fibers, focal adhesions and the ECM, and how such interaction regulates the development of force in the cytoskeletal system.
Toward this end, we have developed the model to be grounded in non-equilibrium thermodynamics, accounting for the mechano-chemistry of focal adhesions, stress fibers and ECM.
This paper's focus is on the development of the model itself and of the theoretical framework for analyzing the mechano-chemistry of such biological systems.
A few simulations results are provided in this communication in order to demonstrate the model at work. A comprehensive study of the model's response needs to be rather extensive because of its complexity, and is beyond the scope of this mainly theoretical paper. Such a study is being provided elsewhere \citep{maraldi:2013}.

\section{An archetypal model for the cytoskeletal system}
\label{sect:model-intro}
\begin{figure}[h]
\centering
\includegraphics[width=0.9\columnwidth]{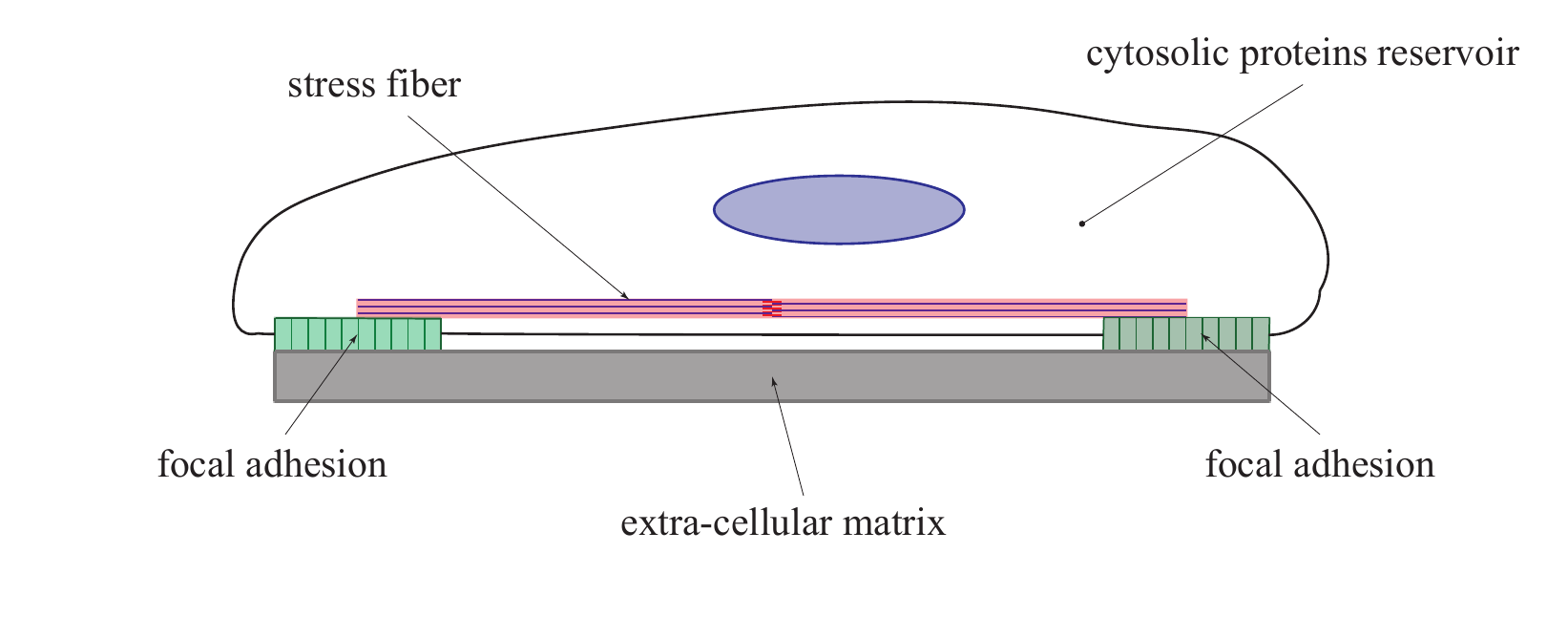}
\caption{\small{Schematic representation of the model.}}
\label{fig:model}
\end{figure}

We consider a minimal system consisting of a stress fiber connected to a focal adhesion at each end, with each focal adhesion being attached to the ECM underlying the cell, as depicted in Fig. \ref{fig:model}; the system also includes the cytosolic reservoir supplying proteins to the stress fiber and the focal adhesions.

We represent the focal adhesions as one-dimensional clusters of generic complexes (each complex being an ensemble of proteins consisting of a ligand, an integrin molecule and associated focal adhesion proteins).
The stress fiber is modeled as a one-dimensional, tension bearing structure composed of a number of generic protein complexes, each incorporating, among others, actin molecules, myosin molecules and ATP molecules to fuel myosin contractility.
The system may be subjected to a mechanical load applied to the ECM in order to reproduce the effects of external perturbations that, in a living cell, may come from a direct action on the ECM itself, from the action of surrounding cells, or from other such stress fiber-focal adhesion systems within the same cell.

It is to be expected that a system formed by coupling two mechano-chemically dependent sub-systems (stress fiber and focal adhesions) will gain in complexity.
In this case, complexity is enhanced beyond the mechano-sensitivity discussed above because of the rate-dependent response of the stress fiber that will be discussed in Sect. \ref{sect:SF-int}.

\subsection{System kinematics resulting from protein addition}
\label{sect:kinematics}
A comprehensive model of the cytoskeletal system has to account for the experimentally observed interplay between mechanical and chemical phenomena; the model proposed here highlights this connection in that protein binding to (or unbinding from) the stress fiber and the focal adhesions results in a change in the geometry of the system, which in turn affects its ability to generate contractile force and to sustain external loads.
It also accounts for the fact that the chemistry of both the stress fiber and the focal adhesions is influenced by the force within the system, as will be explained in the sections to follow.

In general, force is non-uniformly distributed along the focal adhesion.
However, for the sake of simplicity, in the following we will assume a uniform force distribution over the focal adhesion.
This choice allows us to carry out the calculations by considering only the overall force acting on the focal adhesion and the consequent overall deformation.
The hypothesis of uniform force distribution along the focal adhesion is also supported by the results of previous computations \citep{olberding:2010}, which showed that the behavior of the focal adhesion model we adopt in this paper is not significantly affected by the choice of the force distribution. 
In a similar fashion, for the sake of simplicity but while retaining the most relevant physics, we will model the stress fiber as a 1-D subsystem. While accounting for its diameter, the force and deformation will be considered uniformly distributed over the stress fiber's cross section; i.e. the same force in each actin filament.
Hence, again, only the overall force and deformation will matter and the stress fiber will be modeled as a rod-like element.
The number of proteins in the stress fiber, $N_\mathrm{sf}$, is related to its geometry as:
\begin{equation} \label{eq:radius}
r_\mathrm{sf} = \sqrt{ \frac{N_\mathrm{sf} V_\mathrm{act}}{\pi x_\mathrm{sf}^0} },
\end{equation}
where $r_\mathrm{sf}$ is the stress fiber radius, $V_\mathrm{act}$ is the volume of a single actin monomer and $x_\mathrm{sf}^0$ is the unstretched stress fiber length.
Another quantity relevant to the geometry of the system is the number of actin filaments in the stress fiber, which is given by:
\begin{equation}
\label{eq:Nfil}
N_\mathrm{fil} = \frac{N_\mathrm{sf} L_\mathrm{actmon}}{x_\mathrm{sf}^0},
\end{equation}
where $L_\mathrm{actmon}$ is the length of an actin monomer.
Note that as the system evolves and proteins bind to or unbind from the stress fiber, its radius (and, likewise, the number of actin filaments) evolves.

\begin{figure}[h]
\centering
\includegraphics[scale=1.0]{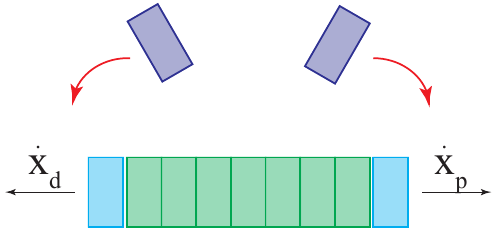}
\caption{F\small{A complex binding/unbinding and resulting focal adhesion ends velocities.}}
\label{fig:fa-bind}
\end{figure}
Each complex in the focal adhesion has length $\lambda$. Complexes can bind and unbind anywhere along the focal adhesion. In our model we track the concentration of these complexes, $c_\mathrm{fa}$, which is a number density per unit length along the focal adhesion. However, we only need to track binding and unbinding at the proximal and distal ends to compute the change in focal adhesion length. This has been depicted in Fig.  \ref{fig:fa-bind}. Accordingly, the focal adhesion ends move at the following velocities \citep{olberding:2010}:
\begin{subequations} \label{eq:edges-speed}
\begin{align}
& \dot{x}_\mathrm{p} =   \lambda^2 \dot{c}_\mathrm{fa}^\mathrm{p} \\
& \dot{x}_\mathrm{d} = - \lambda^2 \dot{c}_\mathrm{fa}^\mathrm{d},
\end{align}
\end{subequations}
where ${x}_\mathrm{p}$ and ${x}_\mathrm{d}$ are the positions of the proximal and distal ends, respectively, and $\dot{c}_\mathrm{fa}^\mathrm{p}$, $\dot{c}_\mathrm{fa}^\mathrm{d}$ are the corresponding rates of change of concentration of focal adhesion complexes due to protein binding/unbinding. Knowing ${x}_\mathrm{p}$ and ${x}_\mathrm{d}$, the focal adhesion length is:
\begin{equation}
\hat{x}_\mathrm{fa} = \vert {x}_\mathrm{p} - {x}_\mathrm{d}\vert.
\end{equation}
When the velocities of the two ends have the same direction, focal adhesion \emph{translation} is observed. It is best quantified as a motion of the focal adhesion's centroid. If $\tilde{x}_\mathrm{fa}$ is the resultant translation of the centroid relative to the reference (initial) position $\tilde{x}_\mathrm{fa} = 0$ at time $t = 0$, we write:
\begin{equation}
\label{eq:fa-cent-transl}
\dot{\tilde{x}}_\mathrm{fa} = \frac{1}{2}\brac{\dot{x}_\mathrm{p} + \dot{x}_\mathrm{d}}
                            = \frac{\lambda^2}{2}\brac{\dot{c}_\mathrm{fa}^\mathrm{p} - \dot{c}_\mathrm{fa}^\mathrm{d}}.
\end{equation}                                          
Focal adhesion translation is likely associated with protein \textit{treadmilling} through the focal adhesion and cytosol.\footnote{In \cite{olberding:2010} the term \emph{treadmilling} has been applied to the mechanism of focal adhesion translation.} As we will see below, focal adhesion translation provides a mechanism for force relaxation within the system.

\begin{figure} [t]
\centering
\includegraphics{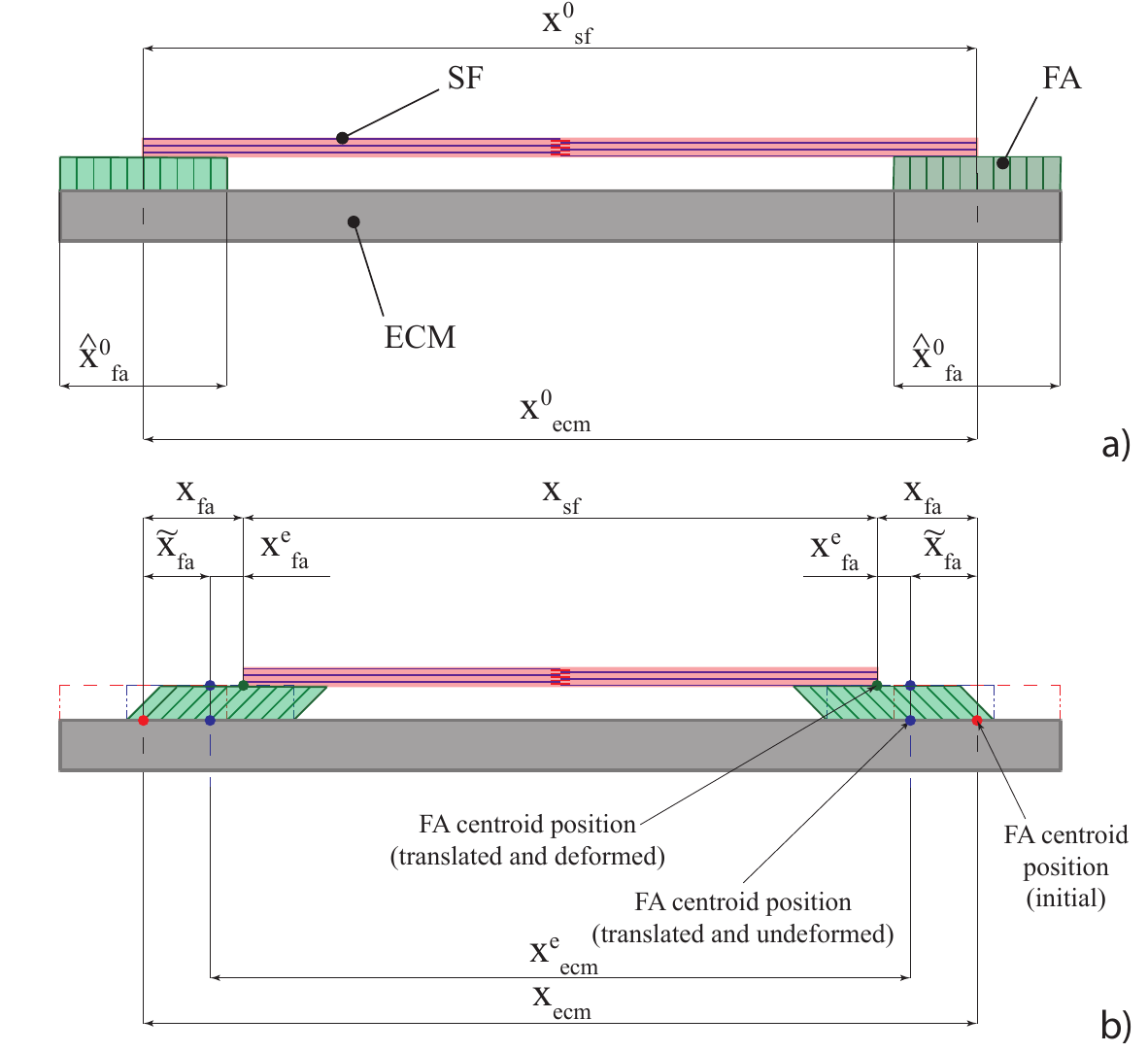}
\caption{\small{Schematic representation of the kinematics of the model. a) the system in the initial, undeformed state; b) the system in the current state, including the effects of translation of the focal adhesions and deformation of both the focal adhesions and the stress fiber. Note that $x_\mathrm{ecm} = x^0_\mathrm{ecm}$.}}
\label{fig:system-len}
\end{figure}
As depicted in Fig. \ref{fig:system-len}, under the action of mechanical load, the focal adhesion deforms elastically as well as translates; therefore, the distance between the current position of the focal adhesion's centroid and its initial position is given by:
\begin{equation}
\label{eq:fa-observable}
x_\mathrm{fa} = x_\mathrm{fa}^\mathrm{e} + \tilde{x}_\mathrm{fa},
\end{equation}
where $x_\mathrm{fa}^\mathrm{e}$ is the magnitude of the centroid's displacement due to elastic deformation of the focal adhesion. For more details on the additive decomposition implied by Eq.\eqref{eq:fa-observable} see Fig. \ref{fig:system-len}b.

\subsection{Stress fiber internal variables}
\label{sect:SF-int}
An accurate description of the mechanical response of the stress fiber has to account for its sarcomeric structure, due to which anti-parallel actin filaments can: (a) actively slide past each other due to the stepping action of myosin molecules and (b) passively slide past each other on account of the tension-dependent breakage of bonds between $\alpha$-actinin and actin. 
These two effects endow the stress fibers with a rate-dependent response, which can be modeled by means of active stress and viscoelastic models, respectively. 
A classical work on the force-dependent kinetics of myosin stepping on actin filaments is that by \citet{hill:1938}, while the viscoelastic response of stress fibers has been demonstrated by \citet{kumar:2006}.

\begin{figure} 
\centering
\includegraphics{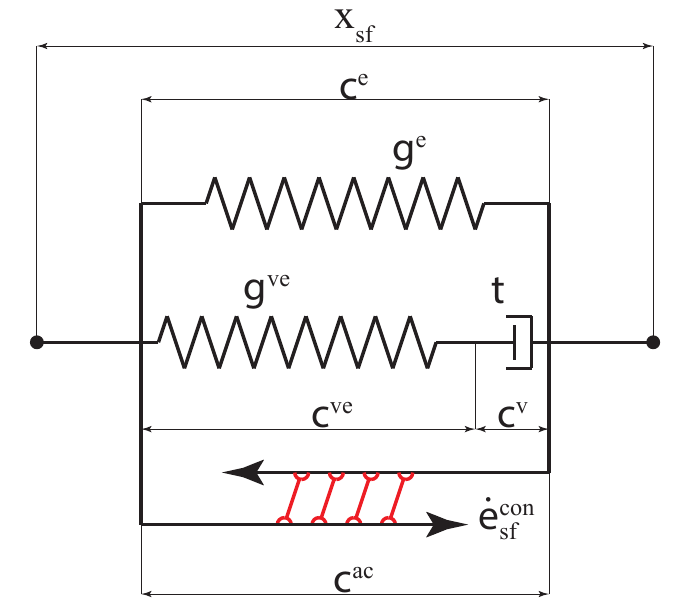}
\caption{\small{Constitutive model for the contractile force in the stress fiber. See the text for definitions of the symbols used.}}
\label{fig:rehological}
\end{figure}

In this treatment we are only concerned with the force at the ends of the stress fiber and its overall deformation (see Sects.\ref{thermo:basis}--\ref{sect:nonequilibrium}).
For this reason, we are not modeling explicitly the individual sarcomeric units; rather, we account for their collective effect.
Following the structural model proposed by \citet{pellegrin:2007}, we represent the stress fiber rheology using a purely elastic element in parallel with a Maxwell viscoelastic element and a contractile element (Fig.  \ref{fig:rehological}).

For a complete description of the state of the stress fiber it is necessary to introduce a number of internal variables, namely the displacement of the purely elastic element $\xi^\mathrm{e}$, the displacement of the elastic part of the Maxwell element $\xi^\mathrm{ve}$, the displacement of the viscous element $\xi^\mathrm{v}$, and the contraction of the active element $\xi^\mathrm{ac}$.
The internal force conjugate to $\xi^\mathrm{e}$ is the elastic force $P^\mathrm{e}_\mathrm{sf}$, that conjugate to $\xi^\mathrm{ve}$ and $\xi^\mathrm{v}$ is the visco-elastic force $P^\mathrm{ve}_\mathrm{sf}$, and that conjugate to $\xi^\mathrm{ac}$ is the active force $P^\mathrm{ac}_\mathrm{sf}$.
According to the schematic representation in Fig.  \ref{fig:rehological}, the stress fiber internal variables can be related to one another and to the current length of the stress fiber $x_\mathrm{sf}$ (the observable variable) as:
\begin{equation}
\label{eq:SF-const-displ}
d x_\mathrm{sf} = d\xi^\mathrm{e} = d\xi^\mathrm{ve} + d\xi^\mathrm{v} = d\xi^\mathrm{ac},
\end{equation}
whereas the force acting on the stress fiber is related to the internal forces through the following equation: 
\begin{equation}
\label{eq:SF-const-force}
P_\mathrm{sf} = P_\mathrm{sf}^\mathrm{e} + P_\mathrm{sf}^\mathrm{ve} + P_\mathrm{sf}^\mathrm{ac}.
\end{equation}
Contraction of the active element happens only upon ATP hydrolysis, which releases ADP; the reaction occurs under the following chemical constraint on the number of ATP and ADP molecules:
\begin{equation}
dN_\mathrm{ATP} + dN_\mathrm{ADP} = 0,
\label{atp-adp-constraint}
\end{equation}
which accounts for the stoichiometry of the ATP-to-ADP conversion reaction.

\section{The thermodynamic basis}
\label{thermo:basis}
In the current Section and in Sect.  \ref{sect:equilibrium}, we review standard equilibrium thermodynamics as applied to the model in order to build the necessary background for the non-equilibrium thermodynamic treatment of Sect. \ref{sect:nonequilibrium}.
In turn, the non-equilibrium treatment will pose restrictions on constitutive equations and rate laws in Sects. \ref{constforcesect}--\ref{sect:rate-laws}.

For modeling purposes, the system is split into four sub-systems: the two focal adhesions, the stress fiber, the cytosol and the ECM.
Mechanics and chemistry are both relevant to the stress fiber and focal adhesion sub-systems; the cytosolic reservoir is a purely chemical sub-system; the ECM is a purely mechanical sub-system. 
For symmetry reasons (Fig. \ref{fig:system-len}), the two focal adhesions are identical in geometry and response.

We consider internal energy functions with the following parameterizations for the different sub-systems:
\begin{equation} \label{intengies}
\begin{aligned}
&U_\mathrm{sf}  = U_\mathrm{sf}  \brac{N_\mathrm{sf}, S_\mathrm{sf}, \xi^\mathrm{e}_\mathrm{sf}, \xi^\mathrm{ve}_\mathrm{sf}, N_\mathrm{ATP}, N_\mathrm{ADP}}\\
&U_\mathrm{fa}  = U_\mathrm{fa}  \brac{x^\mathrm{e}_\mathrm{fa}, N_\mathrm{fa}, S_\mathrm{fa}} \\
&U_\mathrm{cyt} = U_\mathrm{cyt} \brac{\widehat{N}_\mathrm{sf}, \widehat{N}_\mathrm{fa}, S_\mathrm{cyt}} \\
&U_\mathrm{ecm} = U_\mathrm{ecm} \brac{x^\mathrm{e}_\mathrm{ecm}, S_\mathrm{ecm}}
\end{aligned}
\end{equation}
where $U_{\alpha}$ is the internal energy and $S_{\alpha}$ is the entropy of sub-system $\alpha$.
The variables $N_\mathrm{sf}$ and $N_\mathrm{fa}$ are, respectively, the numbers of generic proteins in the stress fiber and of complexes in the focal adhesion, while $\widehat{N}_\mathrm{sf}$ and $\widehat{N}_\mathrm{fa}$ are, respectively, the numbers of generic stress fiber proteins and focal adhesion complexes in the cytosol.
Note that $\xi^\mathrm{e}_\mathrm{sf}$ and $\xi^\mathrm{ve}_\mathrm{sf}$ are the only internal variables responsible for internal energy storage in the stress fiber (in the form of elastic energy). Finally, $x^\mathrm{e}_\mathrm{ecm}$ is the length of the ECM subjected to elastic deformation (see Fig. \ref{fig:system-len} and the explanation below).

Given Eq. \eqref{intengies}, the entropy functions are parametrized as:
\begin{equation}\label{entropies}
\begin{aligned}
& S_\mathrm{sf}  = S_\mathrm{sf}  \brac{N_\mathrm{sf}, U_\mathrm{sf}, \xi^\mathrm{e}, \xi^\mathrm{ve}, N_\mathrm{ATP}, N_\mathrm{ADP}}\\
&S_\mathrm{fa}  = S_\mathrm{fa}  \brac{x^\mathrm{e}_\mathrm{fa}, N_\mathrm{fa}, U_\mathrm{fa}} \\
&S_\mathrm{cyt} = S_\mathrm{cyt} \brac{\widehat{N}_\mathrm{sf}, \widehat{N}_\mathrm{fa}, U_\mathrm{cyt}} \\
&S_\mathrm{ecm} = S_\mathrm{ecm} \brac{x^\mathrm{e}_\mathrm{ecm}, U_\mathrm{ecm}}.
\end{aligned}
\end{equation}
Assuming that the system is adiabatic leads to the following constraint on internal energies:
\begin{equation} \label{eq:int-nrg}
dU = dU_\mathrm{sf} + 2 \ dU_\mathrm{fa} + dU_\mathrm{cyt} + dU_\mathrm{ecm} = P_\mathrm{ext} dx^\mathrm{e}_\mathrm{ecm},
\end{equation}
which means that the system's energy can only be changed by performing mechanical work on the ECM.
As stated in Sect.  \ref{sect:model-intro}, the agents for this mechanical work could be neighboring cells, other stress fiber - focal adhesions systems within the same cell, or other agents acting through the ECM.

The geometry of the system imposes a kinematic constraint on the system's mechanical extensive variables; with reference to Fig. \ref{fig:system-len} we can write:
\begin{equation} \label{eq:kin-constr}
x_\mathrm{sf} + 2 \ x_\mathrm{fa} = x_\mathrm{ecm},
\end{equation}
where $x_\mathrm{ecm}$ is the length of the ECM that initially underlies the stress fiber -- focal adhesions assembly.
Considering Eq. \eqref{eq:fa-observable} and observing that $x_\mathrm{ecm} = x^\mathrm{e}_\mathrm{ecm} + 2\tilde{x}_\mathrm{fa}$, Eq. \eqref{eq:kin-constr} can be reduced to:
\begin{equation} 
x_\mathrm{sf} + x_\mathrm{fa}^\mathrm{e} = x_\mathrm{ecm}^\mathrm{e},
\end{equation}
or, in incremental form:
\begin{equation} \label{eq:kin-constr-dlta}
dx_\mathrm{sf} + 2 \ dx_\mathrm{fa}^\mathrm{e} - dx_\mathrm{ecm}^\mathrm{e} = 0.
\end{equation}

In addition, the following chemical constraints hold, since the total numbers of generic stress fiber and focal adhesion proteins are conserved:
\begin{equation} \label{eq:chem-constr}
\begin{aligned}
\widehat{N}_\mathrm{sf} + N_\mathrm{sf} &= N_\mathrm{sf}^\mathrm{max}, \\
\widehat{N}_\mathrm{fa} + 2 \ N_\mathrm{fa} &= N_\mathrm{fa}^\mathrm{max},
\end{aligned}
\end{equation}
where $N_\mathrm{sf}^\mathrm{max}$ and $N_\mathrm{fa}^\mathrm{max}$ are the maximum numbers of proteins available to the stress fiber and to a single focal adhesion, respectively.
Eq. \eqref{eq:chem-constr} can also be written in incremental form as:

\begin{equation} \label{eq:chem-constr-dlta}
\begin{aligned}
d\widehat{N}_\mathrm{sf} + dN_\mathrm{sf} &= 0, \\
d\widehat{N}_\mathrm{fa} + 2 \ dN_\mathrm{fa} &= 0.
\end{aligned}
\end{equation}

\section{Thermodynamic equilibrium and its applicability to the modeled processes}
\label{sect:equilibrium}
Motivated by the rapid decay of transient heat fluxes across the cell and by the rapid propagation of elastic waves throughout the system, we assume that the system is at thermal and mechanical equilibrium at each instant of its evolution. In this section we also consider equilibrium with respect to chemistry and the stress fiber's internal variables, in order to obtain constitutive relations rigorously from the thermodynamics. Proper non-equilibrium conditions with respect to chemistry and stress fiber internal variables will be discussed in Sect.  \ref{sect:nonequilibrium}. The framework that we lay out in this section has been arrived at by applying classical equilibrium thermodynamics to this system following the steps laid out in \cite{callen:1985}.

Equilibrium is characterized by a maximum of the system entropy \citep{callen:1985}: 
\begin{equation} \label{eq:entropy-max}
dS = dS_\mathrm{sf} + 2 \ dS_\mathrm{fa} + dS_\mathrm{cyt} + dS_\mathrm{ecm}  = 0.
\end{equation}
Differentiating the entropy function of each sub-system with respect to its natural variables listed in Eq. \eqref{entropies} and using the constraints represented by Eq. \eqref{eq:int-nrg}, \eqref{eq:kin-constr-dlta} and \eqref{eq:chem-constr-dlta} (and considering Eq. \eqref{eq:SF-const-displ} for the stress fiber rheology), Eq. \eqref{eq:entropy-max} becomes:
\begin{equation} \label{eq:entr-max-verbose}
\small
\begin{aligned}
dS \!&= \!\brac{\frac{\!\partial S_\mathrm{sf}}{\partial U_\mathrm{sf}} 
         \!-\! \frac{\partial S_\mathrm{ecm}}{\partial U_\mathrm{ecm}}\!} \! dU_\mathrm{sf} + \!
      2 \brac{\frac{\partial S_\mathrm{fa}}{\partial U_\mathrm{fa}}
         - \frac{\partial S_\mathrm{ecm}}{\partial U_\mathrm{ecm}}} \! dU_\mathrm{fa} +
      \!\brac{\!\frac{\partial S_\mathrm{cyt}}{\partial U_\mathrm{cyt}}
         \!-\! \frac{\partial S_\mathrm{ecm}}{\partial U_\mathrm{ecm}}\!} \! dU_\mathrm{cyt} + \\
   &+ \brac{\!\frac{\partial S_\mathrm{sf}}{\partial N_\mathrm{sf}}
         \!-\! \frac{\partial S_\mathrm{cyt}}{\partial \widehat{N}_\mathrm{sf}}\!} \! dN_\mathrm{sf} +
      2 \brac{\frac{\partial S_\mathrm{fa}}{\partial N_\mathrm{fa}}
         - \frac{\partial S_\mathrm{cyt}}{\partial \widehat{N}_\mathrm{fa}}} \! dN_\mathrm{fa} +
      \!\brac{\frac{\partial S_\mathrm{sf}}{\partial N_\mathrm{ATP}}
         - \frac{\partial S_\mathrm{sf}}{\partial N_\mathrm{ADP}}} \! dN_\mathrm{ATP} + \\
   &+ 2 \brac{\frac{\partial S_\mathrm{fa}}{\partial x^\mathrm{e}_\mathrm{fa}}
         + \frac{\partial S_\mathrm{ecm}}{\partial x^\mathrm{e}_\mathrm{ecm}}
         + \frac{\partial S_\mathrm{ecm}}{\partial U_\mathrm{ecm}} P_\mathrm{ext}} \! dx^\mathrm{e}_\mathrm{fa} +
      \!\brac{\frac{\partial S_\mathrm{sf}}{\partial x_\mathrm{sf}}
         + \frac{\partial S_\mathrm{ecm}}{\partial x^\mathrm{e}_\mathrm{ecm}}
         + \frac{\partial S_\mathrm{ecm}}{\partial U_\mathrm{ecm}} P_\mathrm{ext}} \! dx_\mathrm{sf} + \\
   &+ \frac{\partial S_\mathrm{sf}}{\partial \xi^\mathrm{ve}} d\xi^\mathrm{ve} = 0.
\end{aligned}
\end{equation}

The condition expressed in \eqref{eq:entr-max-verbose} has to be satisfied for any generic perturbations of the independent variables of the sub-system entropies; this leads to the condition of thermal equilibrium:
\begin{equation}
\frac{\partial S_\mathrm{sf}}{\partial U_\mathrm{sf}}   = 
\frac{\partial S_\mathrm{fa}}{\partial U_\mathrm{fa}}   = 
\frac{\partial S_\mathrm{cyt}}{\partial U_\mathrm{cyt}} = 
\frac{\partial S_\mathrm{ecm}}{\partial U_\mathrm{ecm}},
\label{eq:therm-eq}
\end{equation}
to the condition of mechanical equilibrium:
\begin{equation}
\frac{\partial S_\mathrm{sf}}{\partial x_\mathrm{sf}} = 
\frac{\partial S_\mathrm{fa}}{\partial x_\mathrm{fa}} =  
- \brac{\frac{\partial S_\mathrm{ecm}}{\partial x^\mathrm{e}_\mathrm{ecm}}
      + \frac{\partial S_\mathrm{ecm}}{\partial U_\mathrm{ecm}} P_\mathrm{ext}},
\label{eq:mech-eq}
\end{equation}
and to the conditions of chemical equilibrium of stress fiber and focal adhesion proteins:
\begin{equation}
\begin{aligned}
\frac{\partial S_\mathrm{sf}}{\partial N_\mathrm{sf}} &= \frac{\partial S_\mathrm{cyt}}{\partial \widehat{N}_\mathrm{sf}} \\
\frac{\partial S_\mathrm{fa}}{\partial N_\mathrm{fa}} &= \frac{\partial S_\mathrm{cyt}}{\partial \widehat{N}_\mathrm{fa}}.
\end{aligned}
\label{eq:chem-eq}
\end{equation}
It also gives the condition of equilibrium with respect to the stress fiber's internal variables:
\begin{equation}
\frac{\partial S_\mathrm{sf}}{\partial \xi^\mathrm{ve}} = 0
\label{eq:int-eq1}
\end{equation}
and the condition for the equilibrium state at which ATP does not undergo hydrolysis:
\begin{equation}
\frac{\partial S_\mathrm{sf}}{\partial N_\mathrm{ATP}} = \frac{\partial S_\mathrm{sf}}{\partial N_\mathrm{ADP}}.
\label{eq:atp-adp-eq}
\end{equation}
We note that at chemical equilibrium no net protein exchange between the cytosol and the focal adhesions occurs; therefore, according to Eq. \eqref{eq:fa-cent-transl}, focal adhesion translation also vanishes ($d\tilde{x}_\mathrm{fa} = 0$).
This is a mechanical manifestation of chemical equilibrium. In a similar manner, if the ATP-to-ADP conversion does not occur, active force cannot be generated in the stress fiber; hence Eq. \eqref{eq:atp-adp-eq} implies, via a constitutive relation:
\begin{equation}
\label{eq:ac-force-eq}
P_\mathrm{sf}^\mathrm{ac} = 0.
\end{equation}

At equilibrium, the fundamental relations of thermodynamics can be stated for the separate sub-systems:
\begin{equation}
\begin{aligned}
&T_\mathrm{sf} dS_\mathrm{sf} = dU_\mathrm{sf} - P_\mathrm{sf}^\mathrm{e} d\xi^\mathrm{e} - P_\mathrm{sf}^\mathrm{ve} d\xi^\mathrm{ve} - \mu_\mathrm{sf} dN_\mathrm{sf} + \brac{\mu_\mathrm{ATP} - \mu_\mathrm{ADP}} dN_\mathrm{ADP} \\
&T_\mathrm{fa} dS_\mathrm{fa} = dU_\mathrm{fa} - P_\mathrm{fa} dx^\mathrm{e}_\mathrm{fa} - \mu_\mathrm{fa} dN_\mathrm{fa} \\
&T_\mathrm{cyt} dS_\mathrm{cyt} = dU_\mathrm{cyt} - \mu^\mathrm{sf}_\mathrm{cyt} d\widehat{N}_\mathrm{sf} - \mu^\mathrm{fa}_\mathrm{cyt} d\widehat{N}_\mathrm{fa} \\
&T_\mathrm{ecm} dS_\mathrm{ecm} = dU_\mathrm{ecm} - P_\mathrm{ecm} dx^\mathrm{e}_\mathrm{ecm},
\end{aligned}
\label{eq:fundrel}
\end{equation}
where $T_\alpha$ is the temperature and $\mu_\alpha$ the chemical potential of the corresponding sub-system $\alpha$, and $\mu_\mathrm{cyt}^\beta$ for $\beta = \mathrm{sf}, \, \mathrm{fa}$, is the chemical potential of the $\beta$ proteins in the cytosol.

By differentiating the entropy functions with respect to their natural variables as expressed in Eq. \eqref{entropies} and comparing them with the fundamental relations \eqref{eq:fundrel}, the equilibrium constitutive relations are obtained:
\begin{equation} \label{eq:const-rel-eq}
\begin{aligned}
&\frac{\partial S_\alpha}{\partial U_\alpha} = \frac{1}{T_\alpha} \\
&\frac{\partial S_\alpha}{\partial N_\alpha} = -\frac{\mu_\alpha}{T_\alpha}, \quad\quad \alpha = \mathrm{sf, fa} \\ 
&\frac{\partial S_\mathrm{cyt}}{\partial \widehat{N}_\alpha} = -\frac{\mu^\mathrm{cyt}_\alpha}{T_\mathrm{cyt}}, \quad \ \alpha = \mathrm{sf, fa} \\
&\frac{\partial S_\alpha}{\partial x^\mathrm{e}_\alpha} = -\frac{P_\alpha}{T_\alpha},\quad\quad \alpha = \mathrm{fa, ecm} \\
&\frac{\partial S_\mathrm{sf}}{\partial \xi^\mathrm{e}} = 
 \frac{\partial S_\mathrm{sf}}{\partial x_\mathrm{sf}} = -\frac{P^\mathrm{e}_\mathrm{sf}}{T_\mathrm{sf}}, \quad \frac{\partial S_\mathrm{sf}}{\partial \xi^\mathrm{ve}} = -\frac{P^\mathrm{ve}_\mathrm{sf}}{T_\mathrm{sf}}\\
&\frac{\partial S_\mathrm{sf}}{\partial N_\mathrm{ATP}} = -\frac{\mu_\mathrm{ATP}}{T_\mathrm{sf}},\quad\quad \ \frac{\partial S_\mathrm{sf}}{\partial N_\mathrm{ADP}} = -\frac{\mu_\mathrm{ADP}}{T_\mathrm{sf}}.
\end{aligned}
\end{equation}
Substituting \eqref{eq:const-rel-eq} in (\ref{eq:therm-eq}--\ref{eq:atp-adp-eq}) and accounting for Eq. \eqref{eq:ac-force-eq} and for the stress fiber rheology defined in Eq. \eqref{eq:SF-const-force}, the equations of equilibrium reduce to:
\begin{subequations}
\begin{align}
&T_\mathrm{sf} = T_\mathrm{fa} = T_\mathrm{cyt} = T_\mathrm{ecm} := T \label{eq:temp-equil} \\
&\mu_\mathrm{sf} = \mu^\mathrm{cyt}_\mathrm{sf},\quad \mu_\mathrm{fa} = \mu^\mathrm{cyt}_\mathrm{fa} \label{eq:chempot-equil}\\
&P_\mathrm{sf} = P_\mathrm{fa} = P_\mathrm{ext} - P_\mathrm{ecm} := P \label{eq:force-equil}\\
&P^\mathrm{ve}_\mathrm{sf} = 0 \label{int-force-equil}\\
&\mu_\mathrm{ATP} = \mu_\mathrm{ADP}. \label{atp-adp-equil}
\end{align}
\end{subequations}

\section{Non-equilibrium processes and restrictions to the model}
\label{sect:nonequilibrium}
We continue to consider thermal and mechanical equilibrium to hold. Therefore, Eq. \eqref{eq:temp-equil} and Eq. \eqref{eq:force-equil} remain valid. On the other hand, we do not seek chemical equilibrium with respect to stress fiber and focal adhesion protein binding, or to ATP-to-ADP conversion, as the kinetics of these phenomena occur on much longer time scales. We also do not seek equilibrium with respect to the mechanical processes internal to the stress fiber.
However, as is commonly done in non-equilibrium thermodynamics we do assume that the constitutive relations (\ref {eq:const-rel-eq}), rigorously obtained at equilibrium, continue to hold far from equilibrium \citep{degrootmazur1984}.
In this setting the entropy rate is:
\begin{equation} \label{eq:entropy-rate}
\dot S = -(\mu_\mathrm{sf} - \mu_\mathrm{cyt}^\mathrm{sf}) \dot{N}_\mathrm{sf} - 2 \ (\mu_\mathrm{fa} - \mu_\mathrm{cyt}^\mathrm{fa}) \dot{N}_\mathrm{fa}
  	   + P_\mathrm{sf}^\mathrm{ve} \dot{\xi}^\mathrm{v} + P_\mathrm{sf}^\mathrm{ac} \dot{x}_\mathrm{sf} + \brac{\mu_\mathrm{ATP} - \mu_\mathrm{ADP}} \dot{N}_\mathrm{ADP}.
\end{equation}
The second law of thermodynamics requires that this rate be non-negative, i.e. $\dot S \ge 0$, hence:
\begin{equation} \label{eq:second-law}
- (\mu_\mathrm{sf} - \mu_\mathrm{cyt}^\mathrm{sf}) \dot{N}_\mathrm{sf} 
- 2 \ (\mu_\mathrm{fa} - \mu_\mathrm{cyt}^\mathrm{fa}) \dot{N}_\mathrm{fa}
+ P_\mathrm{sf}^\mathrm{ve} \dot{\xi}^\mathrm{v}
+ P_\mathrm{sf}^\mathrm{ac} \dot{x}_\mathrm{sf} 
+ \brac{\mu_\mathrm{ATP} - \mu_\mathrm{ADP}} \dot{N}_\mathrm{ADP} \ge 0.
\end{equation}
Compliance with the second law of thermodynamics in the form \eqref{eq:second-law} must be ensured for every admissible process, i.e. for every generic choice of the process vector 
\begin{equation*}
\Gamma = \brac{\dot{N}_\mathrm{sf},\dot{N}_\mathrm{fa},\dot{\xi}^\mathrm{v},\dot{x}_\mathrm{sf},\dot{N}_\mathrm{ADP}}.
\end{equation*}
The corresponding sufficient conditions are that the following relations are satisfied:
\begin{subequations} \label{secondlawsuff}
\begin{align}
&\mathrm{sgn}(\dot{N}_\mathrm{sf}) = \mathrm{sgn}(\mu_\mathrm{cyt}^\mathrm{sf} - \mu_\mathrm{sf}) \label{stress fiber-suff}\\
&\mathrm{sgn}(\dot{N}_\mathrm{fa}) = \mathrm{sgn}(\mu_\mathrm{cyt}^\mathrm{fa} - \mu_\mathrm{fa}) \label{focal adhesion-suff}\\
&\mathrm{sgn}(\dot{N}_\mathrm{ADP}) = \mathrm{sgn}(\mu_\mathrm{ATP} - \mu_\mathrm{ADP}) \label{ATP-suff} \\
&P_\mathrm{sf}^\mathrm{ac} \dot{x}_\mathrm{sf} + \brac{\mu_\mathrm{ATP} - \mu_\mathrm{ADP}} \dot{N}_\mathrm{ADP} \ge 0  \label{active-suff} \\
&P_\mathrm{sf}^\mathrm{ve} \ \dot{\xi}^\mathrm{v} \ge 0.  \label{visc-suff}
\end{align}
\end{subequations}

Eq. \eqref{stress fiber-suff}, Eq. \eqref{focal adhesion-suff} and Eq. \eqref{ATP-suff} follow from considering processes in which, respectively, only $\dot{N}_\mathrm{sf}$, $\dot{N}_\mathrm{fa}$ and $\dot{N}_\mathrm{ADP}$ are non-zero. The same procedure leads to Eq. \eqref{visc-suff}, by considering a particular process in which only $\dot{\xi}^\mathrm{v}$ is non-zero.

We observe that active generation of tension in the stress fiber due to actomyosin contractility implies $P_\mathrm{sf}^\mathrm{ac} \ge 0$ and $\dot{x}_\mathrm{sf} \le 0$.
But this requires $ \brac{\mu_\mathrm{ATP} - \mu_\mathrm{ADP}} \dot{N}_\mathrm{ADP} \ge -P_\mathrm{sf}^\mathrm{ac} \dot{x}_\mathrm{sf}$, a condition that can be satisfied by hydrolysis of ATP to ADP, giving $\dot{N}_\mathrm{ADP} \ge 0$ when $\mu_\mathrm{ATP} \ge \mu_\mathrm{ADP}$. 
The reverse reaction, in which ADP is converted into ATP, does not produce active force; hence, $P_\mathrm{sf}^\mathrm{ac} = 0$ holds and inequality \eqref{active-suff} reduces to Eq. \eqref{ATP-suff}.
Notably, in the case of active force generation, the mechano-chemical coupling requires that $\brac{\mu_\mathrm{ATP} - \mu_\mathrm{ADP}} \dot{N}_\mathrm{ADP}$ exceeds a finite, positive number, not merely that $\brac{\mu_\mathrm{ATP} - \mu_\mathrm{ADP}} \dot{N}_\mathrm{ADP} \ge 0$ (see inequality \eqref{adp-rate}).
This emphasizes the importance of considering the full extent of mechano-chemical coupling via inequality \eqref{active-suff} in satisfying \eqref{eq:second-law}. 

Taken together, the conditions in \eqref{secondlawsuff} represent the restrictions posed by thermodynamics on the constitutive nature of the different sub-systems and on the kinetics of the modeled processes.
In the following sections we will exploit these restrictions to formulate suitable constitutive equations for the mechanical forces, the chemical potentials and for the kinetics of protein binding/unbinding.

\section{Constitutive equations for mechanical forces}
\label{constforcesect}
We begin with constitutive relations for all the components of the stress fiber force appearing in Eq. \eqref{eq:SF-const-force}.
The active component of the stress fiber force is written as:
\begin{equation}
P_\mathrm{sf}^\mathrm{ac} = \frac{P_\mathrm{sf}^\mathrm{stl}}{\dot{\varepsilon}_\mathrm{sf}^\mathrm{con}} \brac{\dot{\varepsilon}_\mathrm{sf}^\mathrm{con} - \frac{\dot{x}_\mathrm{sf}}{x_\mathrm{sf}^0}},
\label{eq:con-force}
\end{equation}
where $\dot{\varepsilon}_\mathrm{sf}^\mathrm{con} < 0$ is the maximum contractile strain rate induced in the stress fiber by myosin molecules stepping along the actin filaments at their maximum velocity, and $P_\mathrm{sf}^\mathrm{stl}$ is the tensile stall force at which the stress fiber cannot contract. This linear force-velocity relationship has been simplified from Hill's original work, which suggested a hyperbolic form \citep{hill:1938}. It also has been extended to allow stress fiber slippage for $P^\mathrm{ac}_\mathrm{sf} > P_\mathrm{sf}^\mathrm{stl}$, as reported by \citet{debold:2005}. 

Noting that the myosin proteins act collectively against a tensile force, we take $P_\mathrm{sf}^\mathrm{stl}$ to be proportional to the number of myosin molecules in the stress fiber, which in turn we assume to be proportional to the number of actin monomers \citep{wu:2005}.
This leads to the relation
\begin{equation}
P_\mathrm{sf}^\mathrm{stl} = P_\mathrm{myos}^\mathrm{stl} \beta N_\mathrm{sf},
\end{equation}
where $P_\mathrm{myos}^\mathrm{stl} $ is the stall force of a single myosin, and $\beta$ is the constant of proportionality relating myosin and actin molecule numbers.
Consistently with this model, we assume that the number of myosin molecules does not affect the maximum contractile strain rate, which is dictated by the stepping velocity of a single myosin molecule:
\begin{equation}
\dot{\varepsilon}_\mathrm{sf}^\mathrm{con} = \frac{\dot{x}_\mathrm{myos}^\mathrm{con}}{x_\mathrm{sf}^0}.
\end{equation}
Under these constitutive assumptions on the stress fiber active force, condition \eqref{active-suff} becomes:
\begin{equation}
\frac{P_\mathrm{myos}^\mathrm{stl} \beta N_\mathrm{sf}}{\dot{x}_\mathrm{myos}^\mathrm{con}} \brac{\dot{x}_\mathrm{myos}^\mathrm{con} - \dot{x}_\mathrm{sf}} \dot{x}_\mathrm{sf}
	+ \brac{\mu_\mathrm{ATP} - \mu_\mathrm{ADP}} \dot{N}_\mathrm{ADP} \ge 0.
\label{active-suff-final}
\end{equation}
Eq. \eqref{active-suff-final} must hold together with \eqref{ATP-suff} for every value of $\dot{x}_\mathrm{sf}$ and $\dot{N}_\mathrm{ADP}$.
This translates into the requirement that:\footnote{Note that the right-hand side of inequality \eqref{adp-rate} is positive, in compliance with condition \eqref{ATP-suff}.}
\begin{equation}
\dot{N}_\mathrm{ADP} \brac{\mu_\mathrm{ATP} - \mu_\mathrm{ADP}} \ge \frac{1}{4} P_\mathrm{myos}^\mathrm{stl} \beta \abs{\dot{x}_\mathrm{myos}^\mathrm{con}} N_\mathrm{sf}.
\label{adp-rate}
\end{equation}
We do not pursue a law for $\dot{N}_\mathrm{ADP}$ here, but assume that \eqref{adp-rate} holds.

For the elastic component of the stress fiber force, we assume the following:
\begin{equation}
P_\mathrm{sf}^\mathrm{e} = \pi r_\mathrm{sf}^2 E_\mathrm{sf} \gamma_\mathrm{e} \brac{\frac{\xi^\mathrm{e}}{x_\mathrm{sf}^0} - 1},
\end{equation}
where $E_\mathrm{sf}$ is the stress fiber's overall Young's modulus and $\gamma_\mathrm{e}$ the non-dimensional elastic modulus for the elastic element in Fig. \ref{fig:rehological}.
Furthermore, satisfaction of condition \eqref{visc-suff} is ensured by the following constitutive relation for the viscous force $P_\mathrm{sf}^\mathrm{ve}$:
\begin{equation}
P_\mathrm{sf}^\mathrm{ve} = \frac{\pi r_\mathrm{sf}^2}{x_\mathrm{sf}^0} \ \eta \ \dot{\xi}^\mathrm{v},
\end{equation}
where $\eta \ge 0$ is the intrinsic viscosity of stress fiber slippage.
For the stress fiber constitutive model outlined in Sect.  \ref{sect:SF-int}, the viscous force also can be expressed as:
\begin{equation} \label{eq:Pve-el}
P_\mathrm{sf}^\mathrm{ve} = \pi r_\mathrm{sf}^2 E_\mathrm{sf} \gamma_\mathrm{ve} \brac{\frac{\xi^\mathrm{ve}}{x_\mathrm{sf}^0} - 1},
\end{equation}
where $\gamma_\mathrm{ve}$ is the non-dimensional elastic modulus for the elastic part of the Maxwell element in Fig. \ref{fig:rehological}.

Finally, the viscous force component also can be re-written as a functional of the observable variable $x_\mathrm{sf}(t)$, via a history-dependent integral, yielding the following expression for the overall stress fiber force: 
\begin{equation}
P_\mathrm{sf} = \frac{P_\mathrm{sf}^\mathrm{stl}}{\dot{\varepsilon}_\mathrm{sf}^\mathrm{con}} \brac{\dot{\varepsilon}_\mathrm{sf}^\mathrm{con} - \frac{\dot{x}_\mathrm{sf}}{x_\mathrm{sf}^0}}
	            + \pi r_\mathrm{sf}^2 E_\mathrm{sf} \gamma_\mathrm{e} \brac{\frac{x_\mathrm{sf}}{x_\mathrm{sf}^0} - 1}
              + \pi r_\mathrm{sf}^2 E_\mathrm{sf} \gamma_\mathrm{ve} \int\limits_{0}^{t} \frac{\dot{x}_\mathrm{sf}\brac{s}}{x_\mathrm{sf}^0} e^{- \brac{t - s}/\tau} \, ds \ ,
\label{eq:mech-const}
\end{equation}
where $\tau = \eta / \brac{E_\mathrm{sf} \gamma_\mathrm{ve}}$ is the viscous relaxation time.
Note that, because we have adopted the standard solid viscoelastic model for the description of the stress fiber viscoelastic behavior, Eq. \eqref{eq:mech-const} can be inverted to yield the stress fiber's current length $x_\mathrm{sf}$ as a function of the force.  
This is unlike the Kelvin-Voigt model used in some studies \citep{besser:2011,besser:2007,kumar:2006,stachowiak:2009}, which results in a non-physical, unbounded force for strain-controlled test cases in which time-discontinuous stretches are applied.

The constitutive equation for the force in the focal adhesion is arrived at by modeling the focal adhesion as a linearly elastic structure:
\begin{equation}
P_\mathrm{fa} =\frac{ \overline{E}_\mathrm{fa} \hat{x}_\mathrm{fa} b}{h} x_\mathrm{fa}^\mathrm{e},
\label{eq:const-fa}
\end{equation}
where $h$ and $b$ are the height and width of the focal adhesion, respectively, and $\overline{E}_\mathrm{fa}$ is the focal adhesion's effective elastic modulus, estimated as follows:
\begin{equation}
\overline{E}_\mathrm{fa} = E_\mathrm{fa} \frac{c_\mathrm{fa}}{c^\mathrm{max}_\mathrm{fa}},
\label{eq:maxEfa}
\end{equation}
where $c^\mathrm{max}_\mathrm{fa} = 1/\lambda$ is the maximum attainable concentration, whereas $E_\mathrm{fa}$ is the elastic modulus of a single protein complex. See \cite{olberding:2010} for details.

The constitutive equation for the force in the ECM also is arrived at by modeling it as a uniform, linearly elastic structure:
\begin{equation}
P_\mathrm{ecm} = K_\mathrm{ecm} \brac{x^\mathrm{e}_\mathrm{ecm} - x_\mathrm{ecm}^0},
 \label{eq:const-ecm}
\end{equation}
where $K_\mathrm{ecm}$ is its stiffness, which depends on the Young's modulus of the ECM material, $E_\mathrm{ecm}$.

\section{Constitutive equations for the chemical potentials}
\label{sect:chem-pots}
The chemical potentials regulating the evolution of the stress fiber are:
\begin{subequations} \label{eq:sf-chem-pots}
\begin{align}
&\mu^\mathrm{sf}_\mathrm{cyt} = H_\mathrm{cyt}^\mathrm{sf} + k_B T \ \mathrm{ln}\bracs{\widehat{N}_\mathrm{sf}/ \brac{N_\mathrm{sf}^\mathrm{max} - \widehat{N}_\mathrm{sf}} } \label{eq:chem-pots_cyt_sf} \\
&\mu_\mathrm{sf} = \frac{1}{2} \frac{\brac{P_\mathrm{sf} x^0_\mathrm{sf}}^2}{E_\mathrm{sf} N_\mathrm{sf}^2 V_\mathrm{act}} + U_\mathrm{sf}^\mathrm{conf} - \frac{P_\mathrm{sf}}{N_\mathrm{fil}} \ d_\mathrm{sf}. \label{eq:chem-pots_sf}
\end{align}
\end{subequations}
The terms on the right-hand side of \eqref{eq:chem-pots_cyt_sf} are respectively, the enthalpy of formation of the generic stress fiber proteins, and their mixing entropy in the cytosol.
We assume here that the stress fiber proteins are compartmentalized in the cytosol and that only a fraction of the total proteins is available to a given stress fiber, proportional to its initial length with constant of proportionality $c_\mathrm{sf}^\mathrm{max}$:
\begin{equation}
N_\mathrm{sf}^\mathrm{max} = c_\mathrm{sf}^\mathrm{max} x_\mathrm{sf}^0.
\end{equation}

In \eqref{eq:chem-pots_sf} the first term on the right-hand side is the strain energy in the stress fiber, the second term is the change in internal energy due to any conformational changes during the binding of stress fiber proteins, and the third term arises due to the work done by the force displacing through $d_\mathrm{sf}$ during the conformational change.

The chemical potentials regulating the evolution of the focal adhesions are:
\begin{subequations} \label{eq:fa-chem-pots}
\small
\begin{align}
&\mu^\mathrm{fa}_\mathrm{cyt} = H_\mathrm{cyt}^\mathrm{fa} + k_B T \ \mathrm{ln}\bracs{\widehat{N}_\mathrm{fa}/ \brac{N_\mathrm{fa}^\mathrm{max} - \widehat{N}_\mathrm{fa}} } \label{eq:chem-pots_cyt_fa}\\
&\mu_\mathrm{fa}^\mathrm{d} = \frac{1}{2} \frac{P_\mathrm{fa}^2 h}{\overline{E}_\mathrm{fa} c_\mathrm{fa} \hat{x}^2_\mathrm{fa} b} \!+\! \frac{1}{2} B \kappa^2 \lambda + U_\mathrm{fa}^{conf} \!-\! P_\mathrm{fa} \!\brac{\!d_\mathrm{fa} \!+\! \frac{\lambda}{2}\!} \label{eq:chem-pots_fa_d} \\
&\mu_\mathrm{fa}^\mathrm{p} = \frac{1}{2} \frac{P_\mathrm{fa}^2 h}{\overline{E}_\mathrm{fa} c_\mathrm{fa} \hat{x}^2_\mathrm{fa} b} \!+\! \frac{1}{2} B \kappa^2 \lambda + U_\mathrm{fa}^{conf} \!-\! P_\mathrm{fa} \!\brac{\!d_\mathrm{fa} \!-\! \frac{\lambda}{2}\!} \label{eq:chem-pots_fa_p}
\end{align}
\end{subequations}
The expression for the chemical potential of focal adhesion proteins differs depending on whether it is evaluated at the distal end or at the proximal end of the focal adhesion, Eq. \eqref{eq:chem-pots_fa_d} and Eq. \eqref{eq:chem-pots_fa_p}, respectively.
The detailed arguments behind the various mechano-chemical contributions to $\mu_\mathrm{cyt}^\mathrm{fa}$, $\mu_\mathrm{fa}^\mathrm{d}$ and $\mu_\mathrm{fa}^\mathrm{p}$ have been laid out in \citet{olberding:2010}. We provide a summary below.

The terms on the right-hand side of \eqref{eq:chem-pots_cyt_fa} are, respectively, the enthalpy of formation and the mixing entropy of the focal adhesion protein complexes in the cytosol.
In Eq. \eqref{eq:chem-pots_fa_d} and Eq. \eqref{eq:chem-pots_fa_p}, the first two terms on the right-hand side are, respectively, due to the strain energy of the focal adhesion and the strain energy involved in the bending of the cell membrane.
In the second term, $B$ is the bending modulus of the cell membrane and $\kappa$ is its curvature. This term arises because the binding and unbinding of focal adhesion proteins can change the local curvature of the cell membrane.
The third term is the change in internal energy due to any conformational changes during the binding of focal adhesion proteins, and the fourth term arises due to the work done by the focal adhesion force displacing through $d_\mathrm{fa}$ during the conformational change. 
Finally, the fifth term arises due to the work done by the force when the focal adhesion center of mass effectively translates by binding/unbinding of proteins at its ends.

\section{Rate laws for the chemical species}
\label{sect:rate-laws}
To complete the model, rate laws for the kinetics of stress fiber and focal adhesion proteins have to be defined.
The following expressions comply with the thermodynamic restrictions posed by Eq. \eqref{stress fiber-suff} and Eq. \eqref{focal adhesion-suff}:

\begin{subequations} \label{eq:chem-evos}
\footnotesize{ 
\begin{align}
\dot{N}_\mathrm{sf} &= \!
  \begin{cases}
    k_\mathrm{sf}^\mathrm{b} \brac{N^\mathrm{max}_\mathrm{sf} \!-\! N_\mathrm{sf}}\! \bracs{\! 1-\exp \! \brac{\frac{\mu_\mathrm{sf}-\mu^\mathrm{sf}_\mathrm{cyt}}{k_B T}}}  \textit{(binding)}\\
    k_\mathrm{sf}^\mathrm{u} \ \exp \brac{\chi_\mathrm{sf}} \bracs{ \exp \brac{-\frac{\mu_\mathrm{sf}-\mu^\mathrm{sf}_\mathrm{cyt}}{k_B T}} -1 } \hspace{0.15cm} \textit{(unbinding)}
  \end{cases} \hspace{-0.5cm}
\label{eq:sf-evo} \\
\dot{c}_\mathrm{fa} &= \!
  \begin{cases}
    k_\mathrm{fa}^\mathrm{b} \bracs{ 1 - \exp \brac{\frac{\mu_\mathrm{fa}-\mu^\mathrm{fa}_\mathrm{cyt}}{k_B T}} } \hspace{1.65cm} \textit{(binding)}\\
    k_\mathrm{fa}^\mathrm{u} \ \exp \brac{\chi_\mathrm{fa}} \bracs{ \exp \brac{-\frac{\mu_\mathrm{fa}-\mu^\mathrm{fa}_\mathrm{cyt}}{k_B T}} - 1} \hspace{0.1cm} \textit{(unbinding) ,}
  \end{cases} \hspace{-0.7cm}
\label{eq:fa-evo}
\end{align}
}
\end{subequations}
where binding occurs for the sub-system $\alpha$ if $\mu_\mathrm{\alpha} - \mu^\mathrm{\alpha}_\mathrm{cyt} \leq 0$, whereas unbinding occurs if $\mu_\mathrm{\alpha} - \mu^\mathrm{\alpha}_\mathrm{cyt} > 0$.
In Eq. \eqref{eq:chem-evos}, $k^\mathrm{b}_\alpha$, $k^\mathrm{u}_\alpha$ are, respectively, the binding and unbinding rates for sub-system $\alpha$, $k_B$ is the Boltzmann constant and $\chi_{\alpha} = \chi_{\alpha} \brac{P}$ is a force-dependent exponent regulating the rapid dissociation of molecular bonds \citep{bell:1978} that is equal to the force-dependent part of the chemical potential for the sub-system $\alpha$.
Eq. \eqref{eq:fa-evo} is evaluated at both the distal and the proximal ends of the focal adhesion and allows for the determination of $\dot{c}_\mathrm{fa}^\mathrm{d}$ and $\dot{c}_\mathrm{fa}^\mathrm{p}$ (Eq. \eqref{eq:edges-speed}).
In \cite{olberding:2010} it was shown that the form of the rate laws in \eqref{eq:fa-evo} can be obtained from Transition State Theory. 
While that derivation was carried out for focal adhesion dynamics, we have adopted the general form here for stress fiber dynamics, also. 

The force-dependent exponents that drive the rapid disassociation of the stress fiber and focal adhesion are:
\begin{subequations} \label{eq:chi}
\begin{align}
\chi_\mathrm{sf} &= \frac{1}{2} \frac{\brac{P_\mathrm{sf} x^0_\mathrm{sf}}^2}{E_\mathrm{sf} N_\mathrm{sf}^2 V_\mathrm{act}}
                 - \frac{P_\mathrm{sf}}{N_\mathrm{fil}} \ d_\mathrm{sf} \label{eq:chi-sf} \\
\chi_\mathrm{fa} &= \begin{cases}
    \frac{1}{2} \frac{P_\mathrm{fa}^2 h}{\overline{E}_\mathrm{fa} c_\mathrm{fa} \hat{x}^2_\mathrm{fa} b} 
      - P_\mathrm{fa} \brac{d_\mathrm{fa} + \frac{\lambda}{2}} \hspace{0.18cm} \textit{(proximal)} \\
    \frac{1}{2} \frac{P_\mathrm{fa}^2 h}{\overline{E}_\mathrm{fa} c_\mathrm{fa} \hat{x}_\mathrm{fa}^2 b} 
      - P_\mathrm{fa} \brac{d_\mathrm{fa} - \frac{\lambda}{2}} \hspace{0.18cm}\textit{(distal).}
  \end{cases} \label{eq:chi-fa}
\end{align}
\end{subequations}

\section{Critical loads for subsystems growth or resorption}
\label{sect:crit-loads}
\begin{figure}[h]
\centering
\includegraphics[scale=0.6]{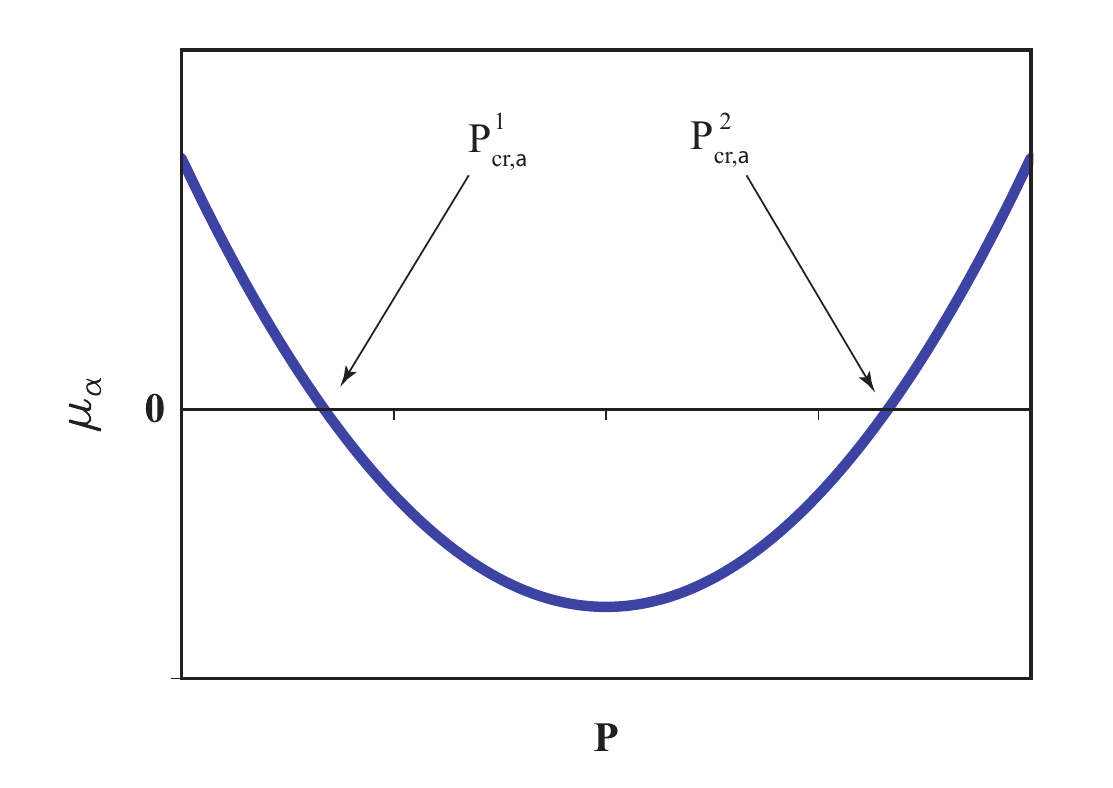}
\caption{\small{Schematic curve of the generic chemical potential $\mu_\alpha$ as a function of the force $P$.}}
\label{fig:chem-pots}
\end{figure}

The mechano-chemical nature of the cytoskeletal system was considered for the development of the model and is manifested in the chemical potentials \eqref{eq:sf-chem-pots} and \eqref{eq:fa-chem-pots} being functions of the force developed within the system (Fig. \ref{fig:chem-pots}).
By comparing the force in the system with suitable critical values it can be established whether a given part of the system (the stress fiber or the focal adhesions) undergoes growth or disassembly.

It is important to recognize, however, that the values of such critical loads vary, as they depend upon the geometry of the system, which is itself subject to change as the system chemically evolves and proteins are recruited to (or disassembled from) the system.

Fig. \ref{fig:chem-pots} is a schematic of the chemical potential of a generic sub-system $\alpha$ (the stress fiber, or the focal adhesion at its distal or proximal end) as a function of the force in the system.
Two critical loads are labeled, which identify regimes over which the sub-system undergoes growth or disassembly.
The specific values of such critical loads for the stress fiber are:
\begin{subequations} \label{eq:P_cr_sf}
\begin{align}
P_\mathrm{cr, sf}^1 &= \frac{N_\mathrm{sf} E_\mathrm{sf} V_\mathrm{act}}{x_\mathrm{sf}^0} 
  \bracs{ \frac{d_\mathrm{sf}}{L_\mathrm{actmon}} - f_\mathrm{sf}\brac{N_\mathrm{sf}} }  \\
P_\mathrm{cr, sf}^2 &= \frac{N_\mathrm{sf} E_\mathrm{sf} V_\mathrm{act}}{x_\mathrm{sf}^0} 
  \bracs{ \frac{d_\mathrm{sf}}{L_\mathrm{actmon}} + f_\mathrm{sf}\brac{N_\mathrm{sf}} },
\end{align}
\end{subequations}
where:
\begin{equation*}
f_\mathrm{sf}\brac{N_\mathrm{sf}} = \sqrt{ \frac{d_\mathrm{sf}^2}{L_\mathrm{actmon}^2} 
- \frac{2 \bracs{U_\mathrm{sf}^\mathrm{conf} - H_\mathrm{cyt}^\mathrm{sf} - k_B T \ ln\brac{\frac{N_\mathrm{sf}^\mathrm{max}}{N_\mathrm{sf}} - 1}}}{E_\mathrm{sf} V_\mathrm{act}} }\ .
\end{equation*}
\vspace{0.2cm}
For the focal adhesion distal end, the critical loads are:
\begin{subequations} \label{eq:P_cr_fad}
\begin{align}
P_\mathrm{cr,fa,d}^1 &= \frac{\hat{x}_\mathrm{fa}^2 b \ \overline{E}_\mathrm{fa} c_\mathrm{fa}}{h} 
  \bracs{ d_\mathrm{fa} - \frac{\lambda}{2} - f_\mathrm{fa,d}\brac{\hat{x}_\mathrm{fa}} }  \\
P_\mathrm{cr,fa,d}^2 &= \frac{\hat{x}_\mathrm{fa}^2 b \ \overline{E}_\mathrm{fa} c_\mathrm{fa}}{h} 
  \bracs{ d_\mathrm{fa} - \frac{\lambda}{2} + f_\mathrm{fa,d}\brac{\hat{x}_\mathrm{fa}} },
\end{align}
\end{subequations}
whereas for the proximal end they are:
\begin{subequations} \label{eq:P_cr_fap}
\begin{align}
P_\mathrm{cr,fa,p}^1 &= \frac{\hat{x}_\mathrm{fa}^2 b \ \overline{E}_\mathrm{fa} c_\mathrm{fa}}{h} 
  \bracs{ d_\mathrm{fa} + \frac{\lambda}{2} - f_\mathrm{fa,p}\brac{\hat{x}_\mathrm{fa}} }  \\
P_\mathrm{cr,fa,p}^2 &= \frac{\hat{x}_\mathrm{fa}^2 b \ \overline{E}_\mathrm{fa} c_\mathrm{fa}}{h}
  \bracs{ d_\mathrm{fa} + \frac{\lambda}{2} + f_\mathrm{fa,p}\brac{\hat{x}_\mathrm{fa}} },
\end{align}
\end{subequations}
where:
\begin{equation*}
f_\mathrm{fa,d}\brac{\hat{x}_\mathrm{fa}} = \sqrt{ d_\mathrm{fa}^2 + \frac{\lambda^2}{4} - \lambda d_\mathrm{fa} 
- \frac{2 h \bracs{U_\mathrm{fa}^\mathrm{conf} + \frac{1}{2} B \kappa^2 \lambda - H_\mathrm{cyt}^\mathrm{fa} - k_B T \ ln\brac{\frac{N_\mathrm{fa}^\mathrm{max}}{N_\mathrm{fa}} - 1}}}{\hat{x}_\mathrm{fa} b \ \overline{E}_\mathrm{fa} c_\mathrm{fa}} }\ ,
\end{equation*}
\begin{equation*}
f_\mathrm{fa,p}\brac{\hat{x}_\mathrm{fa}} = \sqrt{ d_\mathrm{fa}^2 + \frac{\lambda^2}{4} + \lambda d_\mathrm{fa} 
- \frac{2 h \bracs{U_\mathrm{fa}^\mathrm{conf} + \frac{1}{2} B \kappa^2 \lambda - H_\mathrm{cyt}^\mathrm{fa} - k_B T \ ln\brac{\frac{N_\mathrm{fa}^\mathrm{max}}{N_\mathrm{fa}} - 1}}}{\hat{x}_\mathrm{fa} b \ \overline{E}_\mathrm{fa} c_\mathrm{fa}} }\ .
\end{equation*}
\normalsize
\vspace{0.2cm}

If a choice of the model parameters is made such that $0 \le P_\mathrm{cr, \alpha}^1$, the dynamics of the system unfold as follows:
\begin{itemize}
\item [i)] for $P < P_\mathrm{cr, \alpha}^1$, the corresponding chemical potential is positive and the stress fiber undergoes disassembly; however, the associated force-dependent exponent, Eq.  \eqref{eq:chi}, is negative, causing unbinding to be force-penalized.
This regime may be considered as a \textit{slow disassembly} regime for the sub-system;
\item[ii)] for $P_\mathrm{cr, \alpha}^1 < P < P_\mathrm{cr, \alpha}^2$, the corresponding chemical potential is negative and proteins are recruited.
This regime may be considered as a \textit{growth} regime for the sub-system;
\item [iii)] for $P > P_\mathrm{cr, \alpha}^2$, the corresponding chemical potential is positive and the sub-system undergoes disassembly.
The associated force-dependent exponent can be negative or positive; in the latter case, the sub-system undergoes a \textit{force-boosted} (rapid) \textit{disassembly}.
\end{itemize}
The complex mechano-chemical behavior of the cytoskeletal system can thus be modeled as the result of the interplay between the dynamics of all its sub-systems.

\section{The ordinary differential equations governing the evolution of the system}
We now summarize the equations representing the model. Recall that due to the symmetry of the system (Fig. \ref{fig:system-len}), the two focal adhesions are identical in geometry and response, and only one of them needs to be followed.

Our previous computations with a full reaction-diffusion model for focal adhesion dynamics reveal that $c_\mathrm{fa} = c^\mathrm{max}_\mathrm{fa}$ holds almost everywhere, except for a narrow regime immediately behind the proximal edge of a translating focal adhesion.
The condition does hold uniformly for a focal adhesion that is growing at both ends.
Both these results are available as supporting information in \citet{olberding:2010} (see SI Movies 10 and 11).
For this reason, we assume that the focal adhesion elastic modulus attains its maximum value, $\overline{E}_\mathrm{fa} = E_\mathrm{fa}$, corresponding to $c_\mathrm{fa} = c^\mathrm{max}_\mathrm{fa} = 1 / \lambda$ [see Eq. (\ref{eq:maxEfa})].

Under these assumptions, the equations to be solved are:
\begin{subequations} \label{eq:model-ode}
\footnotesize
\begin{align}
\hspace{-0.1cm} \dot{x}_\mathrm{d} &= -\lambda^2 \!
  \begin{cases}
    k_\mathrm{fa}^\mathrm{b} \bracs{ 1 - \exp \brac{\frac{\mu_\mathrm{fa}^\mathrm{d}-\mu^\mathrm{fa}_\mathrm{cyt}}{k_B T}} } \hspace{1.45cm} \textit{(binding)}\\
    k_\mathrm{fa}^\mathrm{u} \ \exp \brac{\chi_\mathrm{fa}} \bracs{ \exp \! \brac{-\frac{\mu_\mathrm{fa}^\mathrm{d}-\mu^\mathrm{fa}_\mathrm{cyt}}{k_B T}} - 1} \hspace{0.0cm} \textit{(unbind.) ,}
  \end{cases} \hspace{-1.1cm} \label{eq:ode-FAd} \\
\hspace{-0.1cm} \dot{x}_\mathrm{p} &= \lambda^2
  \begin{cases}
    k_\mathrm{fa}^\mathrm{b} \bracs{ 1 - \exp \brac{\frac{\mu_\mathrm{fa}^\mathrm{p}-\mu^\mathrm{fa}_\mathrm{cyt}}{k_B T}} } \hspace{1.65cm} \textit{(binding)}\\
    k_\mathrm{fa}^\mathrm{u} \ \exp \brac{\chi_\mathrm{fa}} \! \bracs{ \exp \brac{-\frac{\mu_\mathrm{fa}^\mathrm{p}-\mu^\mathrm{fa}_\mathrm{cyt}}{k_B T}} - 1} \hspace{0.2cm} \textit{(unbind.) ,}
  \end{cases} \hspace{-1.1cm} \label{eq:ode-FAp} \\
\hspace{-0.1cm} \dot{N}_\mathrm{sf} &= \!
  \begin{cases}
    k_\mathrm{sf}^\mathrm{b} \brac{N^\mathrm{max}_\mathrm{sf} \!-\! N_\mathrm{sf}}\! \bracs{ 1 - \exp \! \brac{\frac{\mu_\mathrm{sf}-\mu^\mathrm{sf}_\mathrm{cyt}}{k_B T}}} \hspace{0.4cm} \textit{(binding)}\\
    k_\mathrm{sf}^\mathrm{u} \ \exp \brac{\chi_\mathrm{sf}} \bracs{ \exp \brac{-\frac{\mu_\mathrm{sf}-\mu^\mathrm{sf}_\mathrm{cyt}}{k_B T}} -1 } \hspace{0.6cm} \textit{(unbind.) .}
  \end{cases} \hspace{-0.7cm} \label{eq:ode-SF}
\end{align}
\end{subequations}
The chemical potentials appearing in the system \eqref{eq:model-ode} are, as discussed in Sect. \ref{sect:crit-loads}, functions of the force $P$ within the system, which is provided at every time $t$ by the solution of Eq. \eqref{eq:kin-constr}, Eq. \eqref{eq:force-equil}, Eq. \eqref{eq:mech-const}, Eq. \eqref{eq:const-fa} and Eq. \eqref{eq:const-ecm}.

The system \eqref{eq:model-ode} is constituted of ordinary differential equations and gives as a solution the functions $x^\mathrm{d}(t)$, $x^\mathrm{p}(t)$ and $N_\mathrm{sf}(t)$ (i.e. the time evolution of the FA distal and proximal end positions and of the number of generic stress fiber proteins in the stress fiber) provided that the following initial conditions are specified:
\begin{subequations} \label{eq:model-ic}
\begin{align}
x_\mathrm{d} (0) &= x_\mathrm{d}^0
\label{eq:FAd-ic} \\
x_\mathrm{p} (0) &= x_\mathrm{p}^0
\label{eq:FAp-ic} \\
N_\mathrm{sf} (0) &= \frac{x_\mathrm{sf}^0}{L_\mathrm{actmon}}.
\label{eq:SF-ic}
\end{align}
\end{subequations}

The initial focal adhesion length is therefore,
\begin{equation}
\hat{x}_\mathrm{fa}^0 = \abs{x_\mathrm{p}^0 - x_\mathrm{d}^0}.
\end{equation}
For condition \eqref{eq:SF-ic} we assumed that the stress fiber initially consists of only one filament, in order to model the minimal precursor system, i.e., a single actin filament.

\section{A test case: the influence of the ECM stiffness on cytoskeletal force generation and focal adhesion dynamics}
\label{sect:results}
The model's mechano-chemical response was studied by varying the Young's modulus of the ECM, $E_\mathrm{ecm}$. The parameter values used in the computations are provided as Supporting Information.

For the sake of generality, results are provided in a non-dimensional form.
We chose the set $\bracc{\lambda, E_\mathrm{sf}, k_\mathrm{sf}^\mathrm{b}, c_\mathrm{sf}^\mathrm{max}}$ as the dimensional basis; consequently, $L^\ast = L/\lambda$, $t^\ast = t \ k_\mathrm{sf}^\mathrm{b}$, $P^\ast = P/(\lambda^2 E_\mathrm{sf})$, $N^\ast_\mathrm{sf} = N_\mathrm{sf}/(c_\mathrm{sf}^\mathrm{max} \lambda)$ and $E^\ast = E/E_\mathrm{sf}$ are, respectively, the non-dimensional length, time, force, number of proteins and elastic modulus.

\begin{figure*} [h]
\centering
\includegraphics{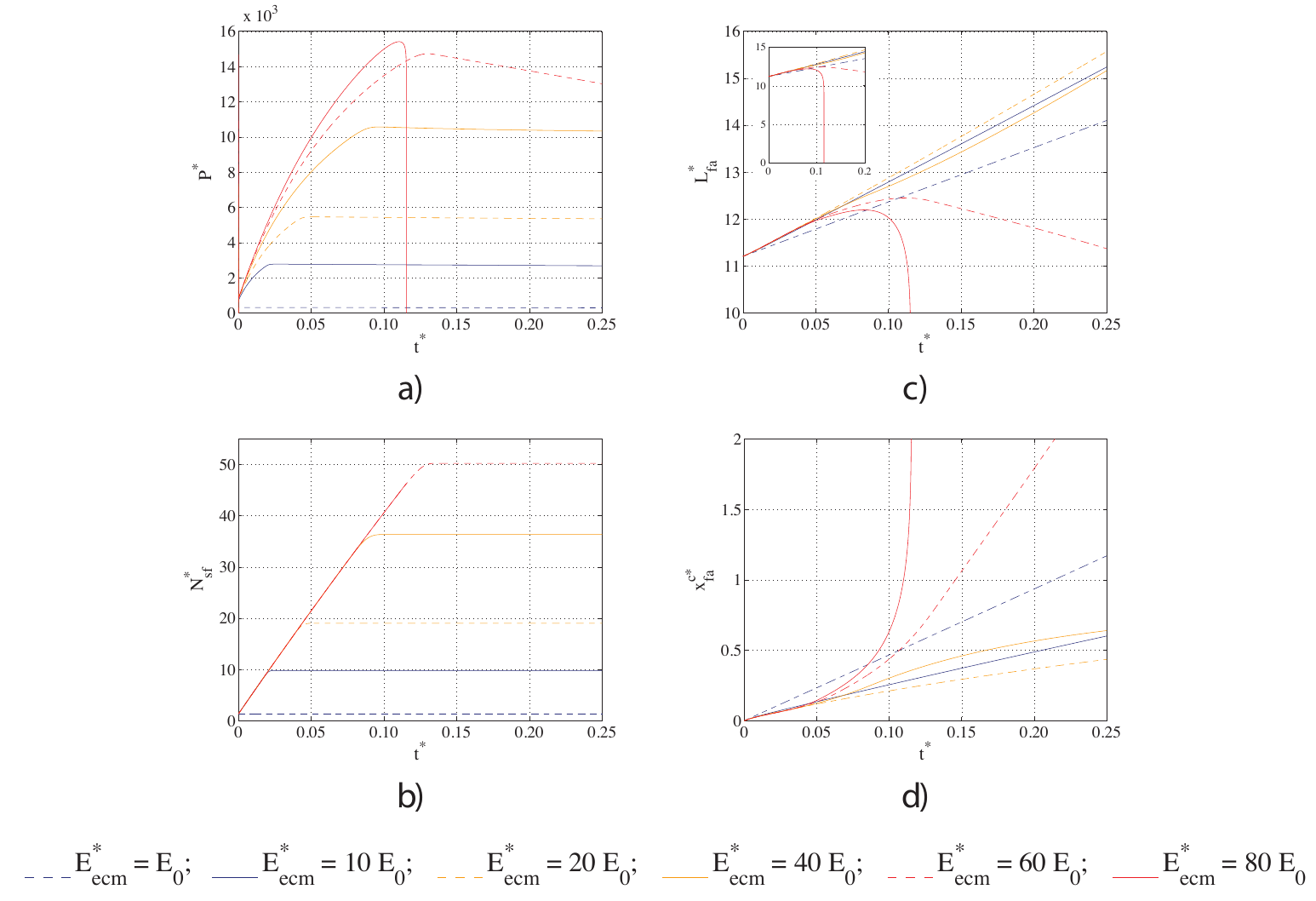}
\caption{\small{Time evolution of: a) Force within the system; b) number of actin monomers in the stress fiber; c) focal adhesion length; d) focal adhesion centroid position for different values of the ECM stiffness. Values are non-dimensional. Non-dimensional reference ECM modulus $E_0 = 6.25 \cdot 10^{-6}$}}
\label{fig:dynamics-un}
\end{figure*}
The detailed dynamics of the system are depicted in Fig. \ref{fig:dynamics-un}.
It can be observed that the force within the system (Fig. \ref{fig:dynamics-un}a) first increases to a maximum value. Then, after a time that depends on the value of the ECM stiffness, it attains a near-plateau with a slightly negative slope.
Our simulations show that increasing the ECM stiffness leads to an increase in the maximum force developed within the system; in this respect, our model provides results in agreement with experimental observations \citep{califano:2010,ghibaudo:2008,kong:2005,lo:2000}.
The force appears to be in direct correlation with the development of the stress fiber: as Fig. \ref{fig:dynamics-un}b shows, a higher ECM stiffness results in more proteins being recruited by the stress fiber.

Pushing the model to its limit, we found that for very stiff ECMs ($E^\ast_\mathrm{ecm} = 80 E_0$) the force drops abruptly after having reached a maximum value, due to the collapse of the focal adhesion. The inset in Fig. \ref{fig:dynamics-un}c shows that in this case the focal adhesion length falls quickly to zero, meaning that the focal adhesion is fully resorbed.
For lower values of $E^\ast_\mathrm{ecm}$, i.e. in the range $E_0$--$20 E_0$, a low value of the stiffness gives rise to focal adhesions that are smaller in size (Fig. \ref{fig:dynamics-un}c) but more motile (as seen in the evolving position of the focal adhesion centroid in Fig. \ref{fig:dynamics-un}d), whereas if the ECM stiffness increases, the focal adhesions become more static and bigger in size.
This trend seems to be in agreement with the experiments reported in literature which show that a compliant substrate promotes more dynamic focal adhesions, as opposed to the more elongated and stable focal adhesions developed in cells sitting on stiff substrates \citep{lo:2000,pelham:1997}.
However, our results do not follow this trend for very high values of the ECM stiffness, as the high value of the force developed within the system under this condition limits the growth of the focal adhesion.
Our simulations also show that a regime of ECM stiffness does exist in which the focal adhesions' size and position do not change dramatically from the initial values.
This is seen for ECM stiffness in the range of $40E_0$--$60E_0$: as depicted in Fig. \ref{fig:dynamics-un}c, the focal adhesion length grows over the time interval considered for ECM stiffness of $40E_0$, while it initially grows, and then begins to decrease, but somewhat gradually for $60E_0$.
Besides, the focal adhesion translation shows evidence of leveling off for ECM stiffness $40E_0$, while it increases for $60E_0$, but at a lower rate than for $80E_0$, (Fig. \ref{fig:dynamics-un}d).
In this regime, the stress fiber-focal adhesion system is not exactly at equilibrium, but the system parameters ($N^\ast_\mathrm{sf}$, $L^\ast_\mathrm{fa}$ and $x^{c^\ast}_\mathrm{fa}$ ) remain within certain bounds over the non-dimensional times considered here.
This could be representative of the typically limited times of $\sim 1$ hour over which the stress fiber-focal adhesion system has been tracked in the experimental literature \citep{riveline:2001,mannetal:2012}. From the currently available experimental literature it is not clear whether the focal adhesions ever attain true equilibrium. Also, it remains difficult, if not impossible, to visualize stress fiber dynamics in live cells, leaving the question of equilibrium open.

\section{Concluding remarks}
The model we have presented is a step toward a thermodynamically consistent treatment of the mechano-chemistry of cytoskeletal force generation, and dynamics.
We note that mechano-chemical interactions dictate the primacy of the thermodynamic treatment to ensure consistency.
In this setting, it is possible to access many aspects of cytoskeletal force generation and dynamics that have been reported in the experimental literature.
While our focus is on the theoretical, non-equilibrium thermodynamic treatment, we also have demonstrated the model's prediction of stress fiber--focal adhesion dynamics on ECMs of varying stiffnesses.

In the model, while the stress fibers do form a stable sub-system, the cytoskeletal system as a whole is never truly at equilibrium, but slowly evolves, mainly due to the dynamic focal adhesions. In discussing the corresponding results in Sect. \ref{sect:results} we have emphasized that experiments have not been carried out for longer times than $\sim 1$ hour, and that over this interval, our model's results are representative of the experimental studies.

While we have noted that the experimental literature currently leaves it unclear whether the stress fiber--focal adhesion system ever attains true equilibrium, we do consider this question in light of the particulars of the computations we have presented here: The non-equilibrium character of the focal adhesions is partly a consequence of a simplifying modeling assumption that we have made: that the force is uniformly distributed over the focal adhesion.
As a result, the condition that leads to a focal adhesion at equilibrium in our model (i.e. both $\mu^\mathrm{p}_\mathrm{fa} = 0$ and $\mu^\mathrm{d}_\mathrm{fa} = 0$) can never be met. 
Exact mechano-chemical equilibrium can be achieved, however, with a non-uniform force distribution in our model.
See \cite{olberding:2010} in this regard. In this case, the maintenance of mechanical equilibrium requires a non-uniform force through the radius of the stress fiber to balance a non-uniform force distribution over the focal adhesion. This can be attained 
by explicitly modeling the single actin filaments, each subjected to a different force.
Separate rate laws of the form \eqref{eq:ode-SF} would be written for each filament.
Besides, as the force acting on one filament would be equal to the force over some interval of the focal adhesion, protein binding/unbinding would be tracked at every point of the focal adhesion itself using \eqref{eq:fa-evo}.
Alternatively, a continuum-level model could be developed in which concentrations of stress fiber and focal adhesion proteins would be tracked, instead of the reduced-order structural models used here.
These developments are beyond the scope of this communication where we have sought to focus on the essential physics with simpler models, but will be presented later.
A more comprehensive investigation on the behavior of the model is the subject of a separate work, where we have focused on specific applications to make further and deeper connection with the experimental literature \citep{maraldi:2013}.

\bibliographystyle{plainnat}      
\bibliography{biblio}

\begin{thebibliography}{48}
\providecommand{\natexlab}[1]{#1}
\providecommand{\url}[1]{\texttt{#1}}
\expandafter\ifx\csname urlstyle\endcsname\relax
  \providecommand{\doi}[1]{doi: #1}\else
  \providecommand{\doi}{doi: \begingroup \urlstyle{rm}\Url}\fi

\bibitem[Arnold et~al.(2004)Arnold, Cavalcanti-Adam, Glass, Blummel, Eck, and
  et~al.]{arnold:2004}
M.~Arnold, E.A. Cavalcanti-Adam, R.~Glass, J.~Blummel, W.~Eck, and et~al.
\newblock Activation of integrin function by nanopatterned adhesive interfaces.
\newblock \emph{ChemPhysChem}, 5:\penalty0 383--388, 2004.

\bibitem[Balaban et~al.(2001)Balaban, Schwarz, Riveline, Goichberg, Tzur,
  Sabanay, and et~al.]{balaban:2001}
N.~Balaban, U.~Schwarz, D.~Riveline, P.~Goichberg, G.~Tzur, I.~Sabanay, and
  et~al.
\newblock Force and focal adhesion assembly: A close relationship studied using
  elastic micropatterned substrates.
\newblock \emph{Nature Cell Biology}, 3:\penalty0 466--473, 2001.

\bibitem[Bell(1978)]{bell:1978}
G.~Bell.
\newblock Models for the specific adhesion of cells to cells.
\newblock \emph{Science}, 200:\penalty0 618--627, 1978.

\bibitem[Besser and Safran(2006)]{besser:2006}
A.~Besser and S.A. Safran.
\newblock Force-induced adsorption and anisotropic growth of focal adhesions.
\newblock \emph{Biophysical Journal}, 90:\penalty0 3469--3484, 2006.

\bibitem[Besser and Schwartz(2007)]{besser:2007}
A.~Besser and U.~Schwartz.
\newblock Coupling biochemistry and mechanics in cell adhesion a model for
  inhomogeneous stress fiber contraction.
\newblock \emph{New Journal of Physics}, 9:\penalty0 425--452, 2007.

\bibitem[Besser et~al.(2011)Besser, Colombelli, Stelzer, and
  Schwartz]{besser:2011}
A.~Besser, J.~Colombelli, E.~Stelzer, and U.~Schwartz.
\newblock Viscoelastic response of contractile filament bundles.
\newblock \emph{Physical Review E}, 83:\penalty0 051902--1--051902--12, 2011.

\bibitem[Califano and Reinhart-King(2010)]{califano:2010}
J.P. Califano and C.A. Reinhart-King.
\newblock Substrate stiffness and cell area predict cellular traction stresses
  in single cells and cells in contact.
\newblock \emph{Cellular and Molecular Bioengineering}, 3:\penalty0 68--75,
  2010.

\bibitem[Callen(1985)]{callen:1985}
H.~Callen.
\newblock \emph{{Thermodynamics and an introduction to thermostatistics}}.
\newblock John Wiley \& Sons, Inc. New York, NY, USA, 1985.

\bibitem[Chan and Odde(2008)]{chan:2008}
C.~Chan and D.~Odde.
\newblock Traction dynamics of filopodia on compliant substrates.
\newblock \emph{Science}, 322:\penalty0 1687--1691, 2008.

\bibitem[Chrzanowska-­‐Wodnicka and Burridge(1996)]{chrzanowska:1996}
M.~Chrzanowska-­‐Wodnicka and K.~Burridge.
\newblock Rho-­‐stimulated contractility drives the formation of stress
  fibers and focal ahesions.
\newblock \emph{Journal of Cell Biology}, 113:\penalty0 1403--1415, 1996.

\bibitem[Colombelli et~al.(2009)Colombelli, Besser, Kress, Reynaud, Girard,
  Caussinus, and et~al.]{colombelli:2009}
J.~Colombelli, A.~Besser, H.~Kress, E.G. Reynaud, P.~Girard, E.~Caussinus, and
  et~al.
\newblock Mechanosensing in actin stress fibers revealed by a close correlation
  between force and protein localization.
\newblock \emph{Journal of Cell Science}, 122:\penalty0 1665--1679, 2009.

\bibitem[de~Groot and Mazur(1984)]{degrootmazur1984}
S.R. de~Groot and P.~Mazur.
\newblock \emph{Nonequilibrium thermodynamics}.
\newblock Dover, 1984.

\bibitem[Debold et~al.(2005)Debold, Patlak, and Warshaw]{debold:2005}
E.~Debold, J.~Patlak, and D.~Warshaw.
\newblock Slip sliding away: Load-­‐dependence of velocity generated by
  skeletal muscle myosin molecules in the laser trap.
\newblock \emph{Biophysical Journal}, 89:\penalty0 L34--L36, 2005.

\bibitem[Deguchi et~al.(2006)Deguchi, Ohashi, and Sato]{deguchi:2006}
S.~Deguchi, T.~Ohashi, and M.~Sato.
\newblock Tensile properties of single stress fibers isolated from cultured
  vascular smooth muscle cells.
\newblock \emph{Journal of Biomechanics}, 39:\penalty0 2603--2610, 2006.

\bibitem[Deshpande et~al.(2006)Deshpande, McMeeking, and Evans]{deshpande:2006}
V.~Deshpande, R.~McMeeking, and A.~Evans.
\newblock A bio-­‐chemo-­‐mechanical model for cell contractility.
\newblock \emph{Proceedings of the National Academy of Sciences}, 103:\penalty0
  14015--­--14020, 2006.

\bibitem[Deshpande et~al.(2008)Deshpande, Mrkisch, McMeeking, and
  Evans]{deshpande:2008}
V.~Deshpande, M.~Mrkisch, R.~McMeeking, and A.~Evans.
\newblock A bio-­‐chemo-­-mechanical model for coupling cell contractility
  with focal adhesion formation.
\newblock \emph{Journal of the Mechanics and Physics of Solids}, 56:\penalty0
  1484--­‐1510, 2008.

\bibitem[Engler et~al.(2006)Engler, Sen, Sweeney, and Discher]{engler:2006}
A.~Engler, S.~Sen, H.~Sweeney, and D.~Discher.
\newblock Matrix elasticity directs stem cell lineage specification.
\newblock \emph{Cell}, 126:\penalty0 677--689, 2006.

\bibitem[Franke et~al.(1984)Franke, Grafe, Schnittler, Seiffge, Mittermayer,
  and Drenckhahn]{franke:1984}
R.~Franke, M.~Grafe, H.~Schnittler, D.~Seiffge, C.~Mittermayer, and
  D.~Drenckhahn.
\newblock Induction of human vascular endothelial stress fibers by fluid shear
  stress.
\newblock \emph{Nature}, 307:\penalty0 648--649, 1984.

\bibitem[Geiger et~al.(2001)Geiger, Bershadsky, Pankov, and
  Yamada]{geiger:2001}
B.~Geiger, A.~Bershadsky, R.~Pankov, and K.~Yamada.
\newblock Transmembrane extracellular matrix-­‐cytoskeleton crosstalk.
\newblock \emph{Nature Reviews in Molecular and Cellular Biology}, 2:\penalty0
  793--805, 2001.

\bibitem[Ghibaudo et~al.(2008)Ghibaudo, Saez, Trichet, Xayaphoummine, Browaeys,
  Silberzan, Buguin, and Ladoux]{ghibaudo:2008}
M.~Ghibaudo, A.~Saez, L.~Trichet, A.~Xayaphoummine, J.~Browaeys, P.~Silberzan,
  A.~Buguin, and B.~Ladoux.
\newblock Traction forces and rigidity sensing regulate cell functions.
\newblock \emph{Soft Matter}, 4:\penalty0 1836--1843, 2008.

\bibitem[Harland et~al.(2011)Harland, Walcott, and Sun]{harland:2011}
B.~Harland, S.~Walcott, and S.~Sun.
\newblock Adhesion dynamics and durotaxis in migrating cells.
\newblock \emph{Physical Biology}, 8:\penalty0
  015011--­‐1--­‐015011--­‐10, 2011.

\bibitem[Hill(1938)]{hill:1938}
A.~Hill.
\newblock The heat of shortening and the dynamics constants of muscle.
\newblock In \emph{Proceedings of the Royal Society of London, Series B},
  volume 126, pages 136--195, 1938.

\bibitem[Hirata et~al.(2008)Hirata, Tatsumi, and Sokabe]{hirata:2008}
H.~Hirata, H.~Tatsumi, and M.~Sokabe.
\newblock Mechanical forces facilitate actin polymerization at focal adhesions
  in a zyxin-independent manner.
\newblock \emph{Journal of Cell Science}, 121:\penalty0 2795--2804, 2008.

\bibitem[Howard(2001)]{howard:2001}
J.~Howard.
\newblock \emph{Mechanics of Motor Proteins and the Cytoskeleton}.
\newblock Sinauer Associates, Sunderland Mass., 2001.

\bibitem[Ingber(2003)]{ingber:2003}
D.~Ingber.
\newblock Tensegrity i. cell structure and hierarchical systems biology.
\newblock \emph{Journal of Cell Science}, 116:\penalty0 1157--1173, 2003.

\bibitem[Kaunas et~al.(2010)Kaunas, Huang, and Hahn]{kaunas:2010}
K.~Kaunas, Z.~Huang, and J.~Hahn.
\newblock A kinematic model coupling stress fiber dynamics with jnk activation
  in response to matrix stretching.
\newblock \emph{Journal of Theoretical Biology}, 264:\penalty0 593--603, 2010.

\bibitem[Kaunas(2008)]{kaunas:2008}
R.~Kaunas.
\newblock {Modeling cellular adaptation to mechanical stress.}
\newblock In \emph{Bioengineering in Cell and Tissue Research}, {Artmann, G.
  and Chien, S. (Ed.)}, pages 317--348. New York:Springer, 2008.

\bibitem[Kaunas et~al.(2005)Kaunas, Nguyen, Usami, and Chien]{kaunas:2005}
R.~Kaunas, P.~Nguyen, S.~Usami, and S.~Chien.
\newblock Cooperative effects of rho and mechanical stretch on stress fiber
  organization.
\newblock \emph{Proceedings of the National Academy of Sciences}, 102:\penalty0
  15895--15900, 2005.

\bibitem[Kong et~al.(2005)Kong, Polte, Alsberg, and Mooney]{kong:2005}
H.J. Kong, T.R. Polte, E.~Alsberg, and D.J. Mooney.
\newblock Fret measurements of cell--traction forces and nano--scale clustering
  of adhesion ligands varied by substrate stiffness.
\newblock \emph{Proceedings of the National Academy of Science}, 102:\penalty0
  4300--4305, 2005.

\bibitem[Kruse and Julicher(2000)]{kruse:2000}
K.~Kruse and F.~Julicher.
\newblock Actively contracting bundles of polar filaments.
\newblock \emph{Physical Review Letters}, 85:\penalty0 1778--­‐1781, 2000.

\bibitem[Kumar et~al.(2006)Kumar, Maxwell, Heisterkamp, Polte, Lele, and
  Salanga]{kumar:2006}
S.~Kumar, I.~Maxwell, A.~Heisterkamp, T.~Polte, T.~Lele, and M.~et~al. Salanga.
\newblock Viscoelastic relaxation of single living stress fibers and its impact
  on cell shape, cytoskeletal organization, and extracellular matrix mechanics.
\newblock \emph{Biophysical Journal}, 90:\penalty0 3762--3773, 2006.

\bibitem[Lo et~al.(2000)Lo, Wang, Dembo, and Wang]{lo:2000}
C-M. Lo, H-B. Wang, M.~Dembo, and Y-I. Wang.
\newblock Cell movement is guided by the rigidity of the substrate.
\newblock \emph{Biophysical Journal}, 79:\penalty0 144--152, 2000.

\bibitem[Lord and Pollard(2004)]{lord:2004}
M.~Lord and T.D. Pollard.
\newblock Ucs protein rng3p activates actin filament gliding by fission yeast
  myosin-ii.
\newblock \emph{Journal of Cell Biology}, 167:\penalty0 315--325, 2004.

\bibitem[Mann et~al.(2012)Mann, Lam, Weng, Sun, and Fu]{mannetal:2012}
J.~Mann, R.~Lam, S.~Weng, Y.~Sun, and J.~Fu.
\newblock A siliconeÐbased stretchable micropost array membrane for monitoring
  liveÐcell subcellular cytoskeletal response.
\newblock \emph{Lab on a Chip}, 12:\penalty0 731--740, 2012.

\bibitem[Maraldi et~al.(2014)Maraldi, Valero, and Garikipati]{maraldi:2013}
M.~Maraldi, C.~Valero, and K.~Garikipati.
\newblock A computational study of stress fiber--focal adhesion dynamics
  governing cell contractility.
\newblock Submitted, 2014.

\bibitem[Olberding et~al.(2010)Olberding, Thouless, Arruda, and
  Garikipati]{olberding:2010}
J.~Olberding, M.~Thouless, E.~Arruda, and K.~Garikipati.
\newblock The non-­‐equilibrium thermodynamics and kinetics of focal
  adhesion dynamics.
\newblock \emph{PLoS One}, 4:\penalty0 e12043, 2010.

\bibitem[Pelham~Jr. and Wang(1997)]{pelham:1997}
R.J. Pelham~Jr. and Y.L. Wang.
\newblock Cell locomotion and focal adhesions are regulated by substrate
  flexibility.
\newblock \emph{Proceedings of the National Academy of Science}, 94:\penalty0
  13661--13665, 1997.

\bibitem[Pellegrin and Mellor(2007)]{pellegrin:2007}
S.~Pellegrin and H.~Mellor.
\newblock Actin stress fibres.
\newblock \emph{Journal of Cell Science}, 120:\penalty0 3491--3499, 2007.

\bibitem[Peterson et~al.(2004)Peterson, Rajfur, Maddox, Freel, Chen, Edlund,
  and et~al.]{peterson:2004}
L.~Peterson, Z.~Rajfur, A.~Maddox, C.~Freel, Y.~Chen, M.~Edlund, and et~al.
\newblock Simultaneous stretching and contraction of stress fibers in vivo.
\newblock \emph{Molecular Biology of the Cell}, 15:\penalty0 3497--3508, 2004.

\bibitem[Pollard and Borisy(2003)]{pollard:2003}
T.~Pollard and G.~Borisy.
\newblock Cellular motility driven by assembly and disassembly of actin
  filaments.
\newblock \emph{Cell}, 112:\penalty0 453--465, 2003.

\bibitem[Riveline et~al.(2001)Riveline, Zamir, Balaban, Schwarz, Ishikazi,
  Narumiya, and et~al.]{riveline:2001}
D.~Riveline, E.~Zamir, N.~Balaban, U.~Schwarz, T.~Ishikazi, S.~Narumiya, and
  et~al.
\newblock Externally applied local mechanical force induces growth of focal
  contacts by an mdia1-­‐dependent and rock-­‐independent mechanism.
\newblock \emph{Journal of Cell Biology}, 153:\penalty0 1175--1185, 2001.

\bibitem[Shemesh et~al.(2005)Shemesh, Geiger, Bershadsky, and
  Kozlov]{shemesh:2005}
T.~Shemesh, B.~Geiger, A.D. Bershadsky, and M.M. Kozlov.
\newblock Focal adhesions as mechanosensors: a physical mechanism.
\newblock \emph{Proceedings of the National Academy of Sciences}, 102:\penalty0
  12383--12388, 2005.

\bibitem[Stachowiak and O'Shaughnessy(2008)]{stachowiak:2008}
M.~Stachowiak and B.~O'Shaughnessy.
\newblock Kinetics of stress fibers.
\newblock \emph{New Journal of Physics}, 10:\penalty0
  025002--­‐1--­‐025002--­‐26, 2008.

\bibitem[Stachowiak and O'Shaughnessy(2009)]{stachowiak:2009}
M.~Stachowiak and B.~O'Shaughnessy.
\newblock Recoil after severing reveals stress fiber contraction mechanisms.
\newblock \emph{Biophysical Journal}, 97:\penalty0 462--471, 2009.

\bibitem[Tan et~al.(2003)Tan, Tien, Pirone, Gray, Bhadriraju, and
  Chen]{tan:2003}
J.~Tan, J.~Tien, D.~Pirone, D.~Gray, K.~Bhadriraju, and C.~Chen.
\newblock Cells lying on a bed of microneedles: an approach to isolate
  mechanical force.
\newblock \emph{Proceedings of the National Academy of Sciences}, 100:\penalty0
  1484--1489, 2003.

\bibitem[Walcott and Sun(2010)]{walcott:2010}
S.~Walcott and S.~Sun.
\newblock A mechanical model of actin stress fiber formation and substrate
  elasticity sensing in adherent cells.
\newblock \emph{Proceedings of the National Academy of Sciences}, 107:\penalty0
  7757--­‐7762, 2010.

\bibitem[Wu and Pollard(2005)]{wu:2005}
J.-­‐Q. Wu and T.~Pollard.
\newblock Counting cytokinesis proteins globally and locally in fission yeast.
\newblock \emph{Science}, 310:\penalty0 310--314, 2005.

\bibitem[Zamir and Geiger(2001)]{zamir:2001}
E.~Zamir and B.~Geiger.
\newblock Molecular complexity and dynamics of cell-­‐matrix adhesions.
\newblock \emph{Journal of Cell Science}, 114:\penalty0 3583--3590, 2001.

\end{thebibliography}
\clearpage
\appendix
\appendixpage
\section{Parameters' values}
\centering
\rotatebox{90}{
\scriptsize
\centering
\rowcolors{1}{}{lightgray}
\begin{tabular} {| r l l >{\scriptsize}l|}
\hline
\bf{Parameter} & \bf{Value} & \bf{Unit} & \bf{Remarks} \\
\hline
\hline
$\hat{x}_\mathrm{fa}^0$ & $3.6 \cdot 10^{-7}$ & m & Initial FA length \\
$\overline{E}_\mathrm{fa}$ & $5.5 \cdot 10^{6}$ & Pa & \parbox{12.5cm}{$= E_\mathrm{fa}$, elastic modulus assumed to be within the percolation limit \cite{maraldi:2013,olberding:2010}. \\ $E_\mathrm{fa}$ estim. for soft, gel-like biological materials} \\
$b$ & $5.0 \cdot 10^{-7}$ & m & Estimate from images in ref. \cite{balaban:2001} and ref. \cite{riveline:2001} \\
$h$ & $1.0 \cdot 10^{-7}$ & m & Rough estimate on the basis of the length of some focal adhesion proteins \cite{zamir:2001} \\
$c_\mathrm{fa}^\mathrm{max}$ & $1.72 \cdot 10^{7}$ & $\mathrm{m^{-1}}$ & $= 1 / \lambda$ \\
$H^\mathrm{fa}_\mathrm{cyt}$ & $0.0$ & $\mathrm{N \cdot m^2}$ & Imposed, focal adhesion does not grow in absence of force \cite{olberding:2010,balaban:2001,riveline:2001} \\
$\kappa$ & $4.0 \cdot 10^{5}$ & $\mathrm{m^{-1}}$ & Estimate from cell height $\sim 5 \mu\mathrm{m}$.\\
$\lambda$ & $5.8 \cdot 10^{-8}$ & m & From ref. \cite{arnold:2004} \\
$U_\mathrm{fa}^\mathrm{conf}$ & $0.0$ & J & Imposed, focal adhesion does not grow in absence of force \cite{olberding:2010,balaban:2001,riveline:2001} \\
$d_\mathrm{fa}$ & $2.9006 \cdot 10^{-8}$ & m & Set to reproduce a variety of observed experimental behaviors \\
$\mu^\mathrm{fa}_\mathrm{cyt}$ & $0.0$ & J & Imposed, focal adhesion does not grow in absence of force \cite{olberding:2010,balaban:2001,riveline:2001} \\
$k_\mathrm{fa}^\mathrm{b}$ & $2.85 \cdot 10^{-3}$ & $\mathrm{s^{-1}}$ & Set to reproduce a variety of observed experimental behaviors \\
$k_\mathrm{fa}^\mathrm{u}$ & $7.98 \cdot 10^{-4}$ & $\mathrm{s^{-1}}$ & Set to reproduce a variety of observed experimental behaviors \\
$x_\mathrm{sf}^0$ & $1.5 \cdot 10^{-5}$ & m & Undeformed SF length \\
$E_\mathrm{sf}$ & $8.0 \cdot 10^{7}$ & Pa & Estimate based on ref. \cite{deguchi:2006} \\
$V_\mathrm{act}$ & $1.047 \cdot 10^{-25}$ & $\mathrm{m^3}$ & Calculated from data on actin length and diameter in ref. \cite{howard:2001} \\
$d_\mathrm{sf}$ & $2.32 \cdot 10^{-9}$ & m & Set to reproduce a variety of observed experimental behaviors \\
$L_\mathrm{actmon}$ & $2.72 \cdot 10^{-9}$ & m & From ref. \cite{howard:2001} \\
$U_\mathrm{sf}^\mathrm{conf}$ & $0.0$ & J & Absorbed into $H_\mathrm{cyt}^\mathrm{sf}$ \\
$H_\mathrm{cyt}^\mathrm{sf}$ & $-2.47 \cdot 10^{-19}$ & J & Set to reproduce a variety of observed experimental behaviors \\
$c_\mathrm{sf}^\mathrm{max}$ & $1.144 \cdot 10^{11}$ & $\mathrm{m^{-1}}$ & \parbox{12.5cm}{Estim. from total number of actin monomers in yeast cytosol \cite{wu:2005}, ratio of volume of cell to yeast cell and considering $50$ SFs with mean length of $10 \mu\mathrm{m}$ in a cell} \\
$k_\mathrm{sf}^\mathrm{b}$ & $2.725 \cdot 10^{-4}$ & $\mathrm{s^{-1}}$ & Adapted from association rate for ATP-actin at the barbed end \cite{pollard:2003} \\
$k_\mathrm{sf}^\mathrm{u}$ & $0.8$ & $\mathrm{s^{-1}}$ & Adapted from association rate for ATP-actin at the pointed end \cite{pollard:2003} \\
$\gamma_\mathrm{e}$ & $0.9$ & & Set to reproduce a variety of observed experimental behaviors \\
$\gamma_\mathrm{ve}$ & $0.1$ & & Set to reproduce a variety of observed experimental behaviors \\
$\tau$ & $10.0$ & s & Estimate from Fig. 3 in ref. \cite{kumar:2006} \\
$\dot{x}_\mathrm{myos}^\mathrm{con}$ & $-5.0 \cdot 10^{-7}$ & $\mathrm{m \cdot s^{-1}}$ & From ref. \cite{lord:2004} \\
$P_\mathrm{myos}^\mathrm{stl}$ & $3.0 \cdot 10^{-11}$ & N & From ref. \cite{wu:2005} \\
$\beta$ & $1.08 \cdot 10^{-3}$ & & From ref. \cite{wu:2005} \\
$A_\mathrm{ecm}$ & $4.0 \cdot 10^{-10}$ & $\mathrm{m^2}$ & ECM cross-sectional area, assuming a 100 $\mathrm{\mu m}$ $\times$ 4 $\mathrm{\mu m}$ rectangular cross-section \\
$k_\mathrm{B}$ & $1.381 \cdot 10^{-23}$ & $\mathrm{J \cdot K^{-1}}$ & Boltzmann's constant \\
$T$ & $310.0$ & K & Test temperature \\
\hline
\end{tabular}
\label{tab:params}
}
\end{document}